\documentclass[structabstract]{aa}  
\usepackage{graphicx}

\usepackage{txfonts}
\usepackage{longtable}
\usepackage{floatflt}
\usepackage{natbib}
\usepackage{lscape}
\usepackage{subfigure}
\usepackage{float}
\bibpunct{(}{)}{;}{a}{}{,}
\begin{document}
   \title{Tidal streams around galaxies in the SDSS DR7 archive}

   \subtitle{I. First results}

   \author{Arpad Miskolczi \inst{}
          \and
          Dominik J. Bomans \inst{}
          \and
          Ralf-J\"urgen Dettmar \inst{}
          }

   \institute{Astronomical Institute of the Ruhr-University Bochum (AIRUB), Universit\"atsstrasse 150, 44801 Bochum\\
              \email{Miskolczi@astro.rub.de, Bomans@astro.rub.de, Dettmar@astro.rub.de}
             }

   \date{\today}

 
  \abstract
   {Models of hierarchical structure formation predict the accretion of smaller satellite galaxies 
onto more massive systems and this process should be accompanied by a disintegration of the smaller companions visible, e.g., in tidal streams.}
   {In order to verify and quantify this scenario we have developed a search strategy for low surface brightness
tidal structures around a sample of 474 galaxies using 
the Sloan Digital Sky Survey DR7 archive. }
   {Calibrated images taken from the SDSS archive were processed in an automated manner and visually inspected for possible tidal streams.}
   {We were able to extract structures at surface brightness levels ranging
from $\sim$ 24 down to ~28 mag $arcsec^{-2}$. A significant number of  tidal streams was found and measured. Their apparent length varies as they seem to be in different stages of accretion. }
   {At least 6\% of the galaxies show distinct stream like features, a total of 19\% show faint features. Several individual cases are described and discussed.}


   \maketitle

\section{Introduction}

Current theories of galactic evolution within the $\Lambda$CDM cosmology predict that galaxies evolved and grew, and are still growing, by accreting small galaxies or even by colliding with large galaxies -  a process known as the hierarchical \textit{bottom-up} model. \citet{bullock2005} derived a model which described the accretion of a small dwarf galaxy into the halo of the host galaxy. Simulations based on that model reproduced many observed properties of Milky Way satellites and their streams. The current theory suggests, that galaxy halos are growing by accretion of small dwarf galaxies, which leave, for limited time, a spur behind as they are tidally disrupted by their host galaxy. Accretion of small galaxies in ongoing minor mergers was detected just in the last decade, mostly in the vicinity of M31 or our own galaxy, the Milky Way (e.g. Ibata et al. 2001b). Here, methods were used like searching for overdensities in the number of stars in our own galaxy to find streams (e.g. \citet{klement2009}). The furthest galaxy which was analyzed by resolving its stars was NGC 891, where \citet{891stream} also found multiple arcs looping around the galaxy. Finding extragalactic streams at galaxies which are to far to be resolved into stars, however, is more complicated. Theoretical simulations show that the expected surface brightness of the streams, depending on their age and intrinsic luminosities, lies between 26 and 40 mag / arcsec$^{⁻2}$ \citep{johnston2008}. This obviously poses a difficulty, since one would require very sensitive observations to achieve such a deep image. Furthermore, the data calibration has to be very accurate. Because of that, most extragalactic streams were discovered by coincidence. \citet{shang1998} reported the discovery of a partial "ring of light" around NGC 5907, which they interpreted as debris from a disrupted dwarf galaxy. This interpretation was later tested by Mart\'{i}nez-Delgado et al. \citep{delgado2008}, by analyzing deeper images of NGC 5907 and comparing them with numerical simulations. They could show that the observed features are consistent with a single minor merger event. \citet{Malin1999} detected several faint features around galaxies, one of which is M104. \citet{sasaki2007} reported on a galaxy threshing system, in which the small threshed galaxy shows two tidal tails, one towards the host galaxy, and one in the opposite direction. Furthermore, \citet{delgado2009} also reported the discovery of yet another spectacular tidal loop associated with NGC 4013, and even more discoveries of tidal streams were presented by \citet{delgado2010}. Finding more extragalactic tidal systems like this will allow us to derive statistical properties of such features and their neighborhood and help fine tune models of galaxy formation and thus helping us to understand the long term evolution of galaxies.

\section{Acquisition and processing of data}
\subsection{Generating an input sample}
In order to find streams more efficiently, some criteria for the input sample are very useful. We used the following criteria:
\begin{itemize}
 \item Inclination. Should be as close to edge-on as possible.
 \item Size. Galaxies should be at least two arcminutes in diameter.
 \item Morphology. Every galaxy type except elliptical galaxies.
\end{itemize}

The reason for these criteria are the following. Since the probability of finding a stream is the highest for an edge-on galaxy \citep{johnston01}, an ideal sample would only contain nearly edge-on galaxies. Because we are using data from the Sloan Digital Sky Survey \citep{sdss}, Data Release 7 \citep{dr7}, with a sampling of 0.396''/pixel \citep{gunncam1998} and need to resolve structures from stars, the galaxies should have some extent to be resolved. Although there are known faint features around elliptical galaxies in clusters, like in the Virgo Cluster \citep{Janowiecki}, \citep{mihos2005}, we excluded them because of the different environment, potentially different physics and the higher relative velocities in clusters.

While analyzing the bulge types of galaxies, \citet{luetticke2000} created a sample of galaxies. In order to be able to analyze the bulge, the inclination of the galaxy has to be at least around 75 $^\circ$ \citep{shaw1990}. Two more selection criteria besides the inclination of the galaxy for their sample were the size of the galaxies with a lower limit of 2' in diameter and the morphology of the galaxies. Only disc galaxies (with a morphological type code of $-3.5 < T < 9.5$) were chosen. The galaxies were chosen from the RC3 catalog \citep{rc3}. We found that the sample created by \citet{luetticke2000} employed the same selection criteria, so we used it as our input sample. After checking which galaxies of this sample are covered by the Sloan Digital Sky Survey DR7 footprint, our input sample was created. Out of 1350 galaxies, there was a match for 474 galaxies.
 
\subsection{Acquisition}
The next step was the acquisition of data. Since the system response of the imaging camera has the highest values in the g', r' and i' filter \citep{gunncam1998}, we chose to get only the corrected frames in these bands. One imaged field covers 13.5 x 9.8 arcminutes, so most of the galaxies lie within one frame. Several objects lie at the border of one field, and because we looked for possibly extended features, we chose a larger FOV around the objects, typically using 6 to 9 fields per object and filter. The header supplemented calibrated files were then downloaded. 
\subsection{Processing data}
\label{processdata}
The processing was done in 5 steps, listed and explained below.
\begin{itemize}
 \item Mosaicing of the fields.
 \item Stacking g', r' and i' to improve SNR.
 \item Subtraction of stars.
 \item Gaussian filtering.
 \item Creation of and division by an artificial flatfield image.
\end{itemize}

The mosaics were created in each filter, using the Montage software package\footnote{see http://montage.ipac.caltech.edu/}. The mosaics of each filter were then stacked using the IRAF task imcombine, in order to improve the SNR of the image. Because the first "detection" is done visually, and a field crowded by stars can confuse the eye, the stacks were processed by SExtractor \citep{Sextractor1996}. In order to get rid of just the stars, but not the galaxy or background information, the source extraction was done in two steps, a method known as ``hot and cold run`` \citep{Caldwell2008}. Step 1 extracted all sources where an area of at least 5 pixels had a value of 1.5$\sigma$. SExtractor now saved a FITS file, containing all detected sources, which of course included the galaxy, because it is obviously larger than 5 pixels. In the second run, the minimal area was set to 800 pixels. In this run, only the main galaxy and some smaller galaxies were detected, and again saved in a FITS file. The final object file was now created, by subtracting the second run file from the first run file. This now left an object file, in which only small sources like stars were present. Using the IRAF task \textit{imarith}, the sources were subtracted from the original stack. To make faint features visible, as a next step we used a circularly symmetric Gaussian filter. There are other possible filter types available in IRAF, like \textit{adaptive} and \textit{hfilter}, both based on the Haar-Transform \citep{fritze77}. Both are able to show faint features, but through testing the different filters with different settings we found that they do not yield better results. Since the Gaussian filter is faster than the adaptive or the hfilter and yields the same results, we decided to use it instead of the other filters. We found by experimenting with the parameters of the Gaussian filter, that with $\sigma = 7$ \citep{daophotpackage} we achieve a good enhancement of faint extended features. This already yielded faint objects, but in order to improve detection, we sometimes needed to create artificial flatfield images. It turned out that mosaics of SDSS images do not always have a flat background - due to bright stars in the vicinity, changing conditions when different stripes were imaged on different nights, and different sensitivities of the CCD sensors in the imaging array - so we used SExtractor again, but this time, to extract the background. The background was extracted, using a mesh size of 512 to find large gradients. This image was now saved, and, using imarith again, the already processed image was divided by this background image.
\subsection{Analysis}
Before we can accurately search for faint features, we need to be aware of possible difficulties of that task.
\subsubsection{Noise}
\begin{figure}[h]
\centering
\includegraphics[height=\columnwidth,angle=90]{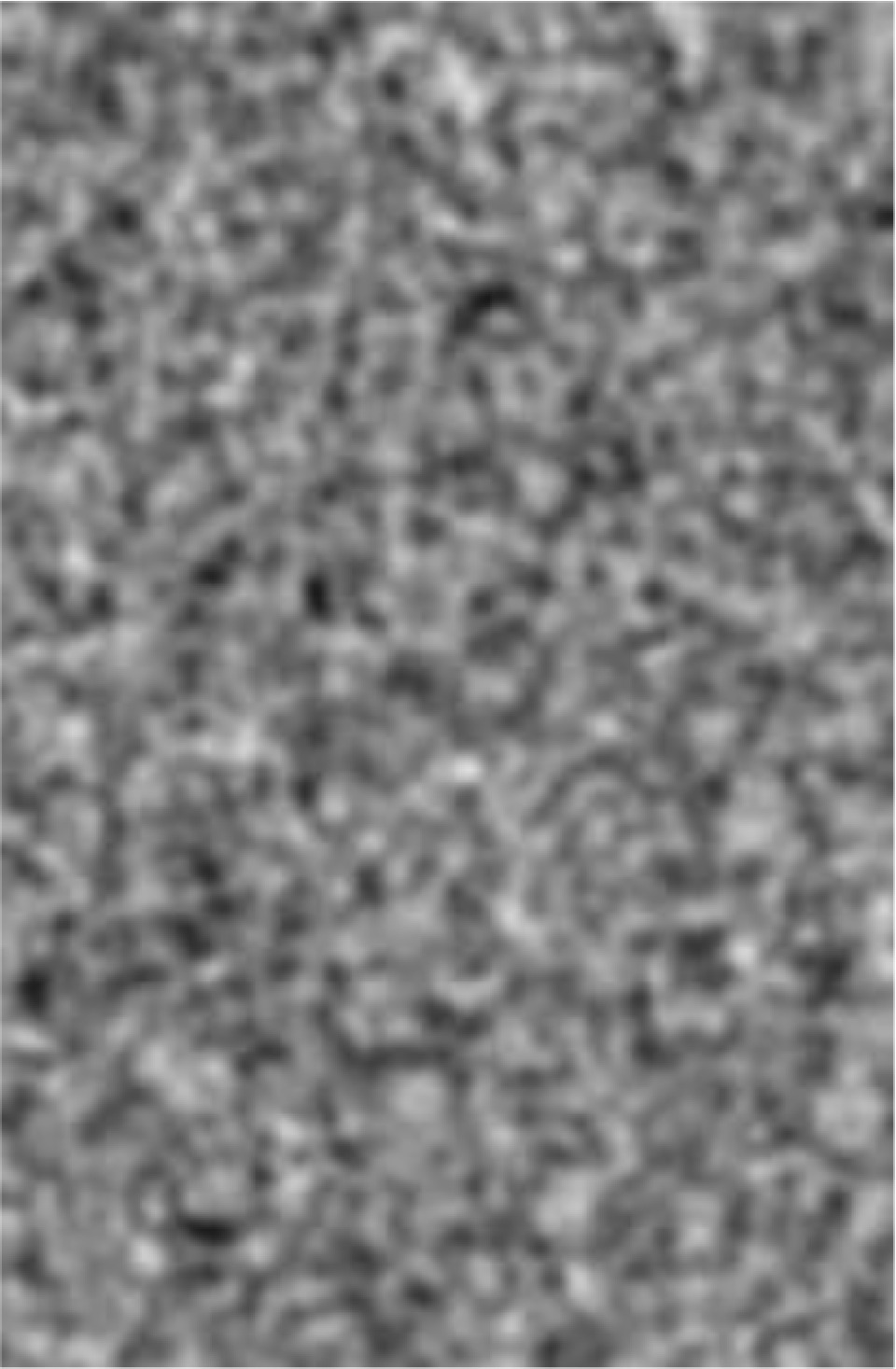}
\caption{Artificial high contrast noise created by IRAF using gain values from SDSS Fits header information, and processed by a Gaussian filter}
\label{noise1}
\end{figure}

The obvious cause of false detections is noise. In order to compare pure noise with our images, we created artificial images using the IRAF task \textsl{artnoise} and instrumental values extracted from the SDSS FITS headers. The images had 2048x2048 pixels and were processed the same way that the original images were, though no sources were extracted since there are none. The artificial noise images showed several "features" which resembled features we saw in the real images. The most common "features" were several pixels thin and had lightning-like shapes, demonstrated in Fig. \ref{noise1}.

Other features we "recreated" were small sources which had quite a value difference to their neighborhood, or in one case, a larger, more diffuse feature. Though these feature could easily be detected as a feature, an argument against them, at least in the science images, is that the faint streams we are looking for have to be associated somehow with a host galaxy. If there is such a "feature" which is just somewhere in the image, not even pointing to a galaxy, we did not regard it as a real feature.

\subsubsection{Residual starlight}
Another effect that can be very misleading is residual starlight. Stars do not show a Gaussian light profile, but rather have additional faint wings \citep{moffat1969}. When they are subtracted from the original image, some very faint parts of them can survive the subtraction, since SExtractor does not consider their wings as part of a star. Individual stars may not contribute much in the resulting image, but when the image has many stars, either in a small group, or seemingly put in a line, the residual light may look like a faint stream, or blob, in the processed image.
\begin{figure}[ht]
\centering
\includegraphics[width=0.8\columnwidth]{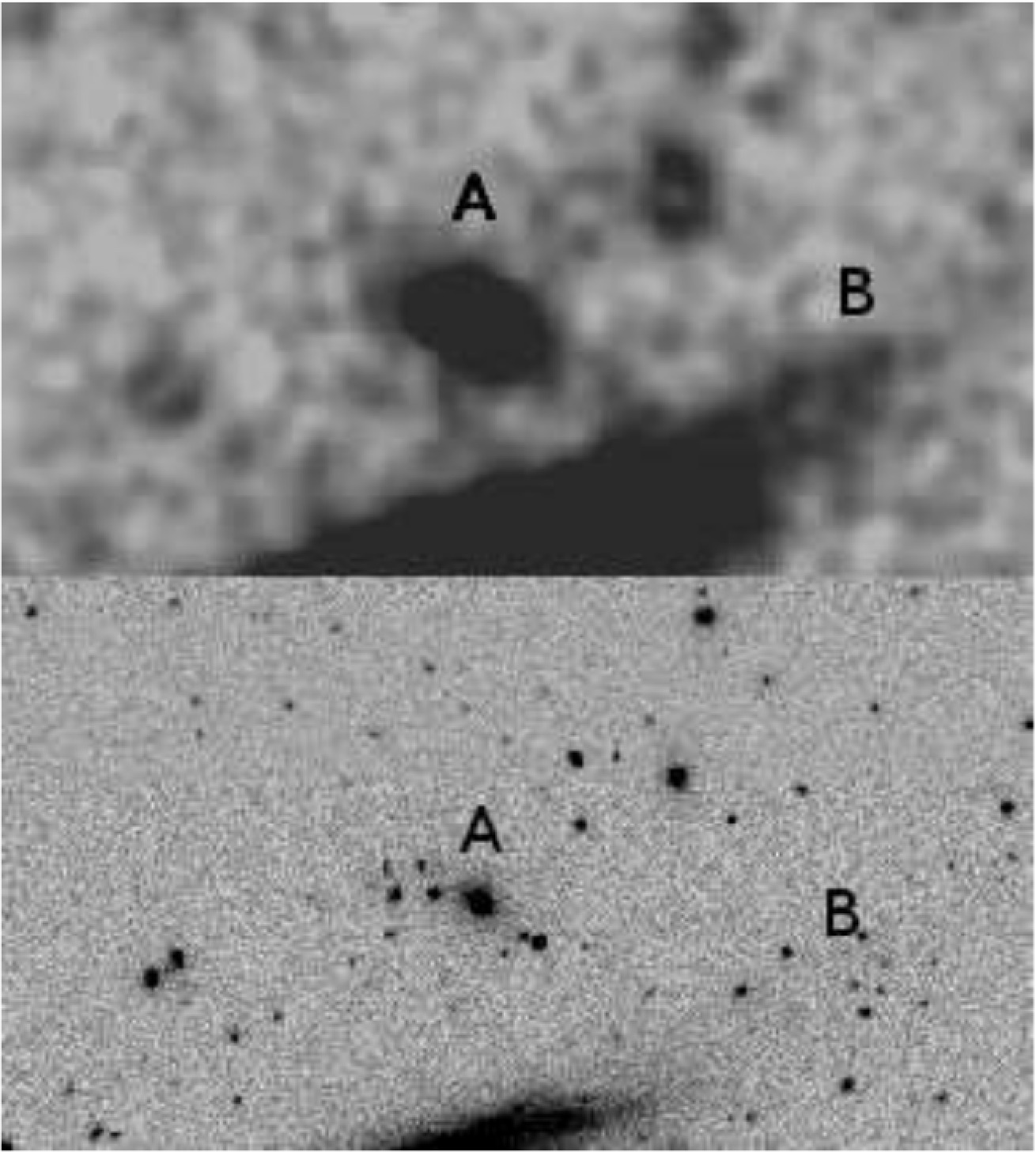}
\caption{Demonstration of effects of residual starlight near UGC 8085. The residual light of the stars is enlarged by the Gaussian filter and can be misleading.}
\label{reslight}
\end{figure}

Fig. \ref{reslight} demonstrates the effects of residual starlight. The stars at position A create features which resemble small streams, emanating from the small galaxy on the left. B is even more misleading, since many stars contribute and make it appear as if there would be a large diffuse feature emanating from the galaxy. Since this is one of the main reasons for possible false detections, the visual inspection of the images was always done by comparing the processed image with its unprocessed counterpart.

\subsection{Is it working?}
Before we even started processing 474 galaxies automatically, we chose to test the method on some galaxies, which have known tidal streams. Our test objects were NGC 5907, NGC 3628, NGC 4013 and NGC 4594, better known as M104, the "Sombrero Galaxy". In addition, a previously unknown stream was observed with the goal of verifying its existence.
\subsubsection{NGC 5907}
\begin{figure}[ht]
\centering
\includegraphics[width=0.9\columnwidth]{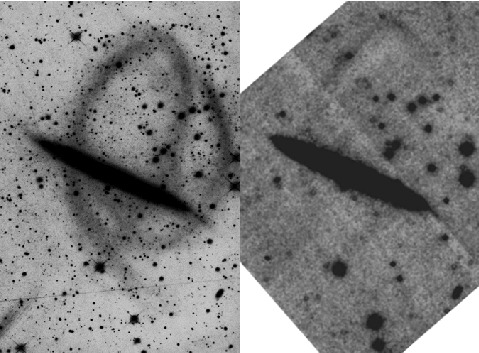}
\caption{Left: The stream around NGC 5907 as observed by \citep{delgado2008}. It is clearly visible. Right: Our processed image of NGC 5907 obtained from the SDSS data with a FOV of 21 by 27 arcminutes. The stream is visible, although it is very faint.}
\label{5907comp}
\end{figure}

After processing the SDSS data of NGC 5907, we had an image which showed that the method is working, see Fig. \ref{5907comp} for a direct comparison between the deep image from Mart\'{i}nez-Delgado et al. and our own.

The processed image obtained from the SDSS data shows that some parts of the NGC 5907 tidal stream are detected. Those parts are the long northern arc on the left and, even fainter, the southern arcs below NGC 5907. Curiously, one of the brightest parts, the arc near the west tip of the galaxy, is not seen in our image. This is due to the afore mentioned effects like different CCD sensitivities which degrade the image quality at low flux counts.

\subsubsection{M104}
In order to test our method in the vicinity of a bright galactic halo, we chose the next target to be M104.
\begin{figure}[ht]
\centering
\includegraphics[width=0.9\columnwidth]{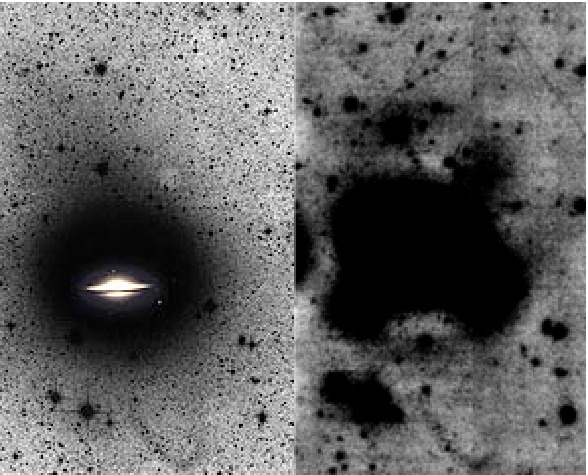}
\caption{Left: Deep image of M104 obtained by David Malin, combining multiple plates taken with the UK Schmidt Telescope. Right: Our image obtained from the SDSS data. Both show a FOV of 22 by 38 arcminutes. The stream can clearly be seen}
\label{m104comp}
\end{figure}
Here, the previously known stream was also detected by our method. Still there is much confusing light in the image, like light from the halos of bright stars, and even from the bright halo of the galaxy itself.
\newpage

\subsubsection{NGC 3628}
As a next test target, we chose the well known tidal Tail of NGC 3628. Once again, we see the feature in the processed SDSS data.
\begin{figure}[ht]
\centering
\includegraphics[width=0.9\columnwidth]{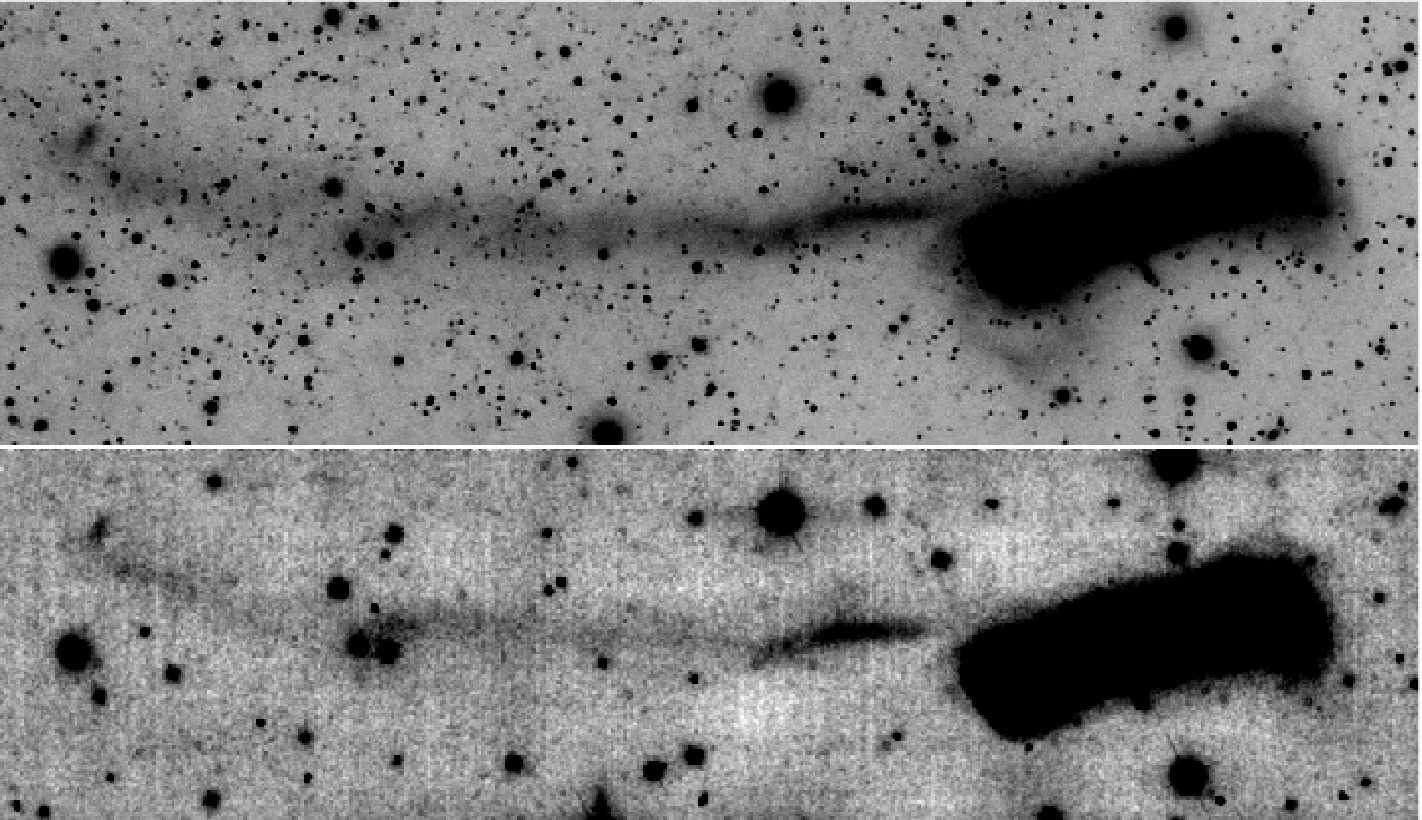}
\caption{Top: Deep image of NGC 3628 with an exposure time of 20 hours, obtained by Steve Mandel. Bottom: Our image obtained from the SDSS data. Both show a FOV of 60 by 15 arcminutes. The tidal tail is very clearly visible}
\label{3628comp}
\end{figure}

Figure \ref{3628comp} shows a direct comparison between a deep image, made by Steve Mandel\footnote{http://www.galaxyimages.com}, using a 6.3" refractor telescope. In this case, the entire feature seems to be visible, though not as spatially detailed as in the comparison  image. This is caused by noise, since these images still have a rather low SNR.

\subsubsection{NGC 4013}
\begin{figure}[ht]
\centering
\includegraphics[width=0.9\columnwidth]{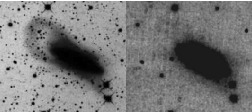}
\caption{Right: Our processed image of NGC 4013 obtained by the SDSS data, with a FOV of 10 by 8 arcminutes. The stream is not visible, although there is a hint of something faint. Left: The stream around NGC 4013 as observed by \citep{delgado2009}}
\label{4013comp}
\end{figure}

As a fourth and last test, we chose NGC 4013 which shows a tidal stream \citep{delgado2009}. As the comparison in Fig. \ref{4013comp} shows, the stream is not convincingly visible in our data. There is a hint of something faint, but if one would not know that there indeed is a faint feature, it is unlikely that one would visually detect it in the SDSS data. Because it has a similar surface brightness as the stream at NGC 5907 \citep{delgado2008}, the missing feature implies that some of the afore mentioned effects lead to a decrease in image quality.

This example shows that though the images go deep, they sometimes suffer from degrading effects which attenuate existing streams. But statistically speaking, our method had a success rate of \textgreater 75 \% on already known streams, which was enough to let it process a large sample. 

\subsection{Verification from other data sources}
In order to know if our method is successful on previously unknown streams, we used other data sources to verify the existence of some of the found features.
\subsubsection{NGC 3509}
In order to test the method and verify the stream at NGC 3509, we observed it with the MONET North remote observatory \citep{Monet}, located near Fort Williams, Texas, USA. 62 images were taken, each with an exposure time of 120 seconds. After every ten images, the pointing was shifted by several pixels for dithering purposes. After flatfield, bias, and darkframe calibrations, the images were stacked using the sigma clipping method of IRAFs \textsl{Imcombine} task. Finally, a Gaussian filter was applied and the contrast severely enhanced.
\begin{figure}[ht]
\centering
\includegraphics[width=0.9\columnwidth]{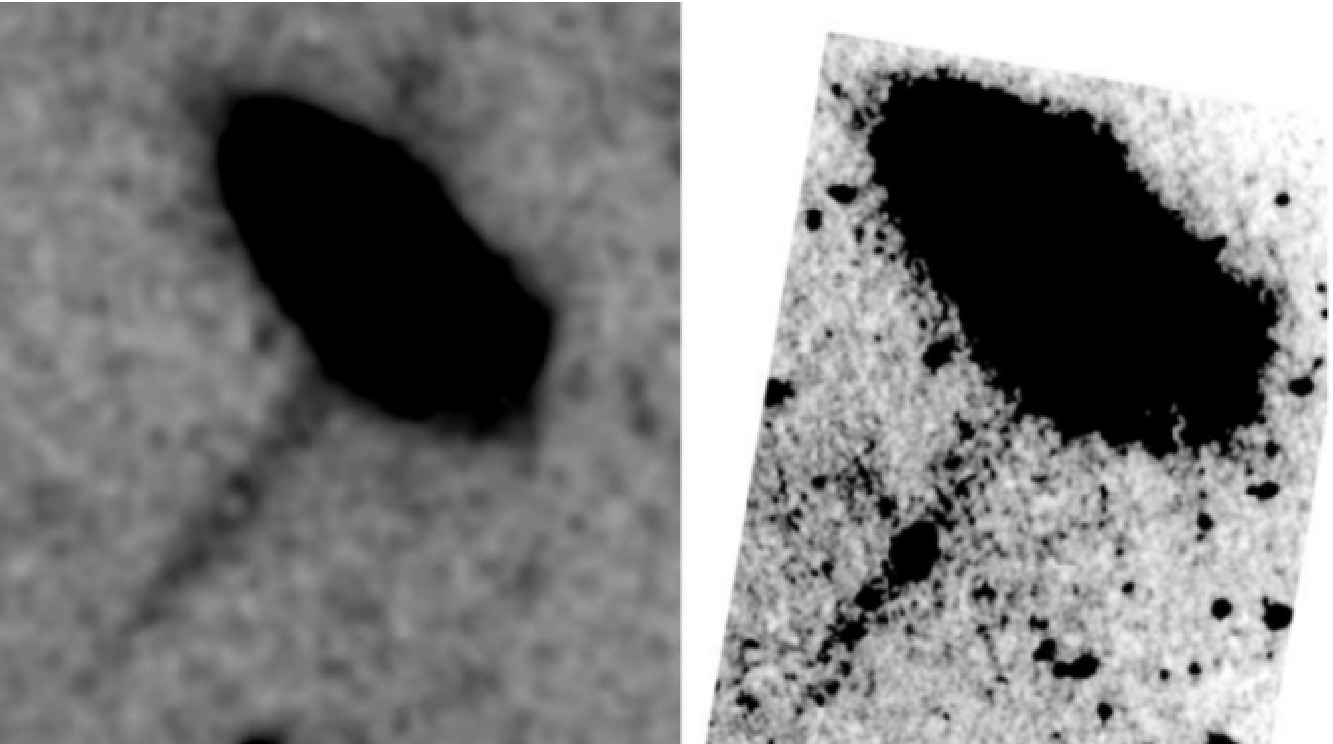}
\caption{Left: Our processed image of NGC 3509 obtained from the SDSS data. Right: Verification of the stream using the Monet NORTH Observatory. Both show a FOV of 4 by 3 arcminutes}
\label{3509comp}
\end{figure}

~\\

There were several observations done with the Hubble Space Telescope which imaged some galaxies of our resulting sample by coincidence, which show a verification of the streams. Those galaxies are NGC 2874 (proposal ID 6357), PGC 1421380 (proposal ID 9765), which lies just below NGC 4206, and NGC 6239 (proposal ID 6359). Those images were processed by applying a Gaussian blur and enhancing the contrast.

\subsubsection{NGC 2874}
Although the feature at NGC 2874 is not considered to be a class I feature (see section \ref{label_res} for our definition of the feature classes), because of the obvious proximity to NGC 2872, it still can serve as a test subject to verify the method.
\begin{figure}[ht]
\centering
\includegraphics[width=\columnwidth]{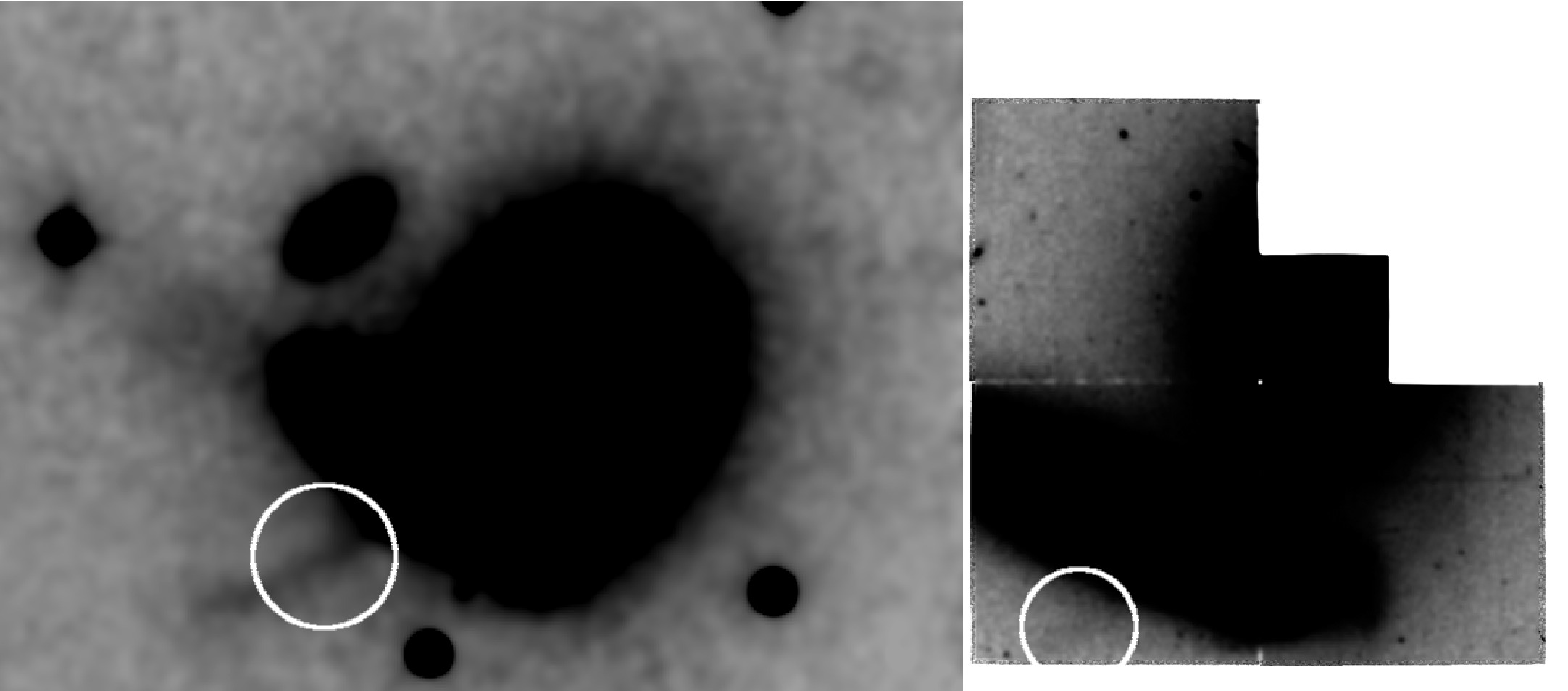}
\caption{Left: Our processed image of NGC 2874 obtained from the SDSS data. It shows a FOV of 4 by 6 arcminutes Right: Processed version of an image of NGC 2874 taken with the Hubble Space Telescope. The stream can be seen in both images.}
\label{2874comp}
\end{figure}

Unfortunately the FOV of the HST image is too small to see the entire stream, but as shown in the marked region of Fig. \ref{2874comp}, a part of the stream is still visible. 

\subsubsection{PGC 1421380}

\begin{figure}[ht]
\centering
\includegraphics[width=0.9\columnwidth]{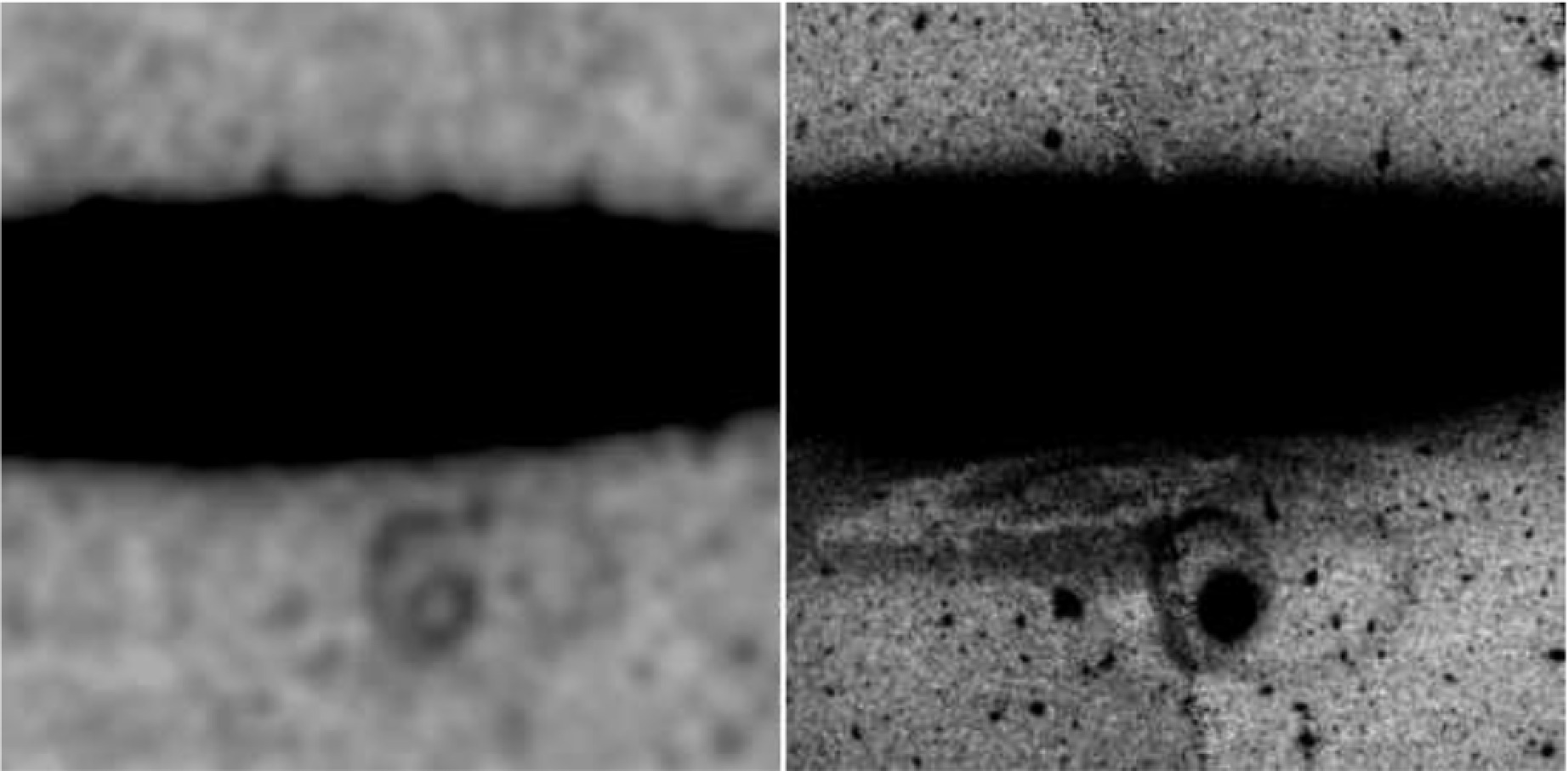}
\caption{Left: Our processed image of PGC 1421380 obtained from the SDSS data. Right: processed version of an image of the same galaxy taken with the Hubble Space Telescope. Both show a FOV of 3 by 3 arcminutes. The stream seems to change its direction in the HST image.}
\label{PGC1421380comp}
\end{figure}

Another curious feature we found in the MAST database is the feature at PGC 1421380, shown in Fig. \ref{PGC1421380comp}, which lies projected below NGC 4206.

This stream like feature surrounding PGC 1421380 is also visible in the HST image, although the preview images are not perfectly calibrated. Curiously, in the SDSS data, the stream looks like it starts at the bottom of the galaxy and streams clockwise up to the small source, but in the HST image, it looks like it starts on top of the galaxy and streams counterclockwise, not even connecting to the small source. The reason for this is probably the lower sampling of the SDSS image, which happens to be in an area, where noise diminishes the SNR. But nonetheless, the stream is considered verified.

\subsubsection{NGC 6239}
\begin{figure}[ht]
\centering
\includegraphics[width=0.9\columnwidth]{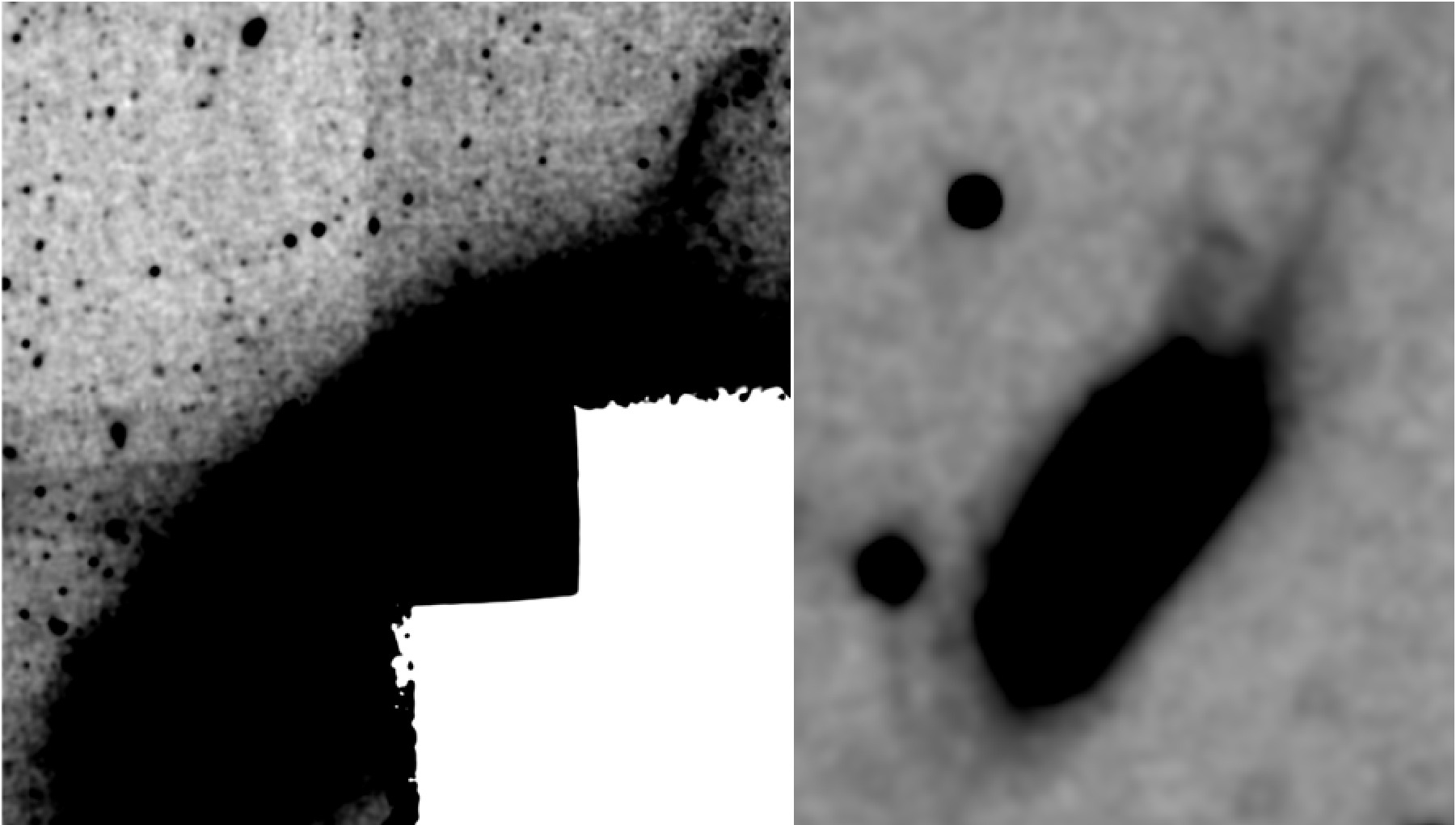}
\caption{Left: Our processed image of NGC 6239 obtained from the SDSS data. It shows a FOV of 4 by 5 arcminutes. Right: Processed version of an image of the same galaxy taken with the Hubble Space Telescope. Only the left feature is visible in the HST image due to the FOV.}
\label{6239comp}
\end{figure}

At NGC 6239, seen in Fig. \ref{6239comp}, the FOV cuts of part of the image, which has the consequence, that we only see one of the two streams. But again, this is enough to convince us of its existence.

\section{Results}
\label{label_res}
After visually inspecting all images, we were left with 126 positive results. Out of those, only 91 were from the input sample and 35 additional features were found at galaxies by coincidence because they happened to be in the same field of view in which the actual target was. The features that we found, were classified in four different classes. Figure \ref{classfreq} shows the distribution of the feature classes, and Figure 11 shows a composition of our most striking results.
\begin{figure*}
\label{bestofcomp}
\centering
\fbox{
\includegraphics[width=\textwidth]{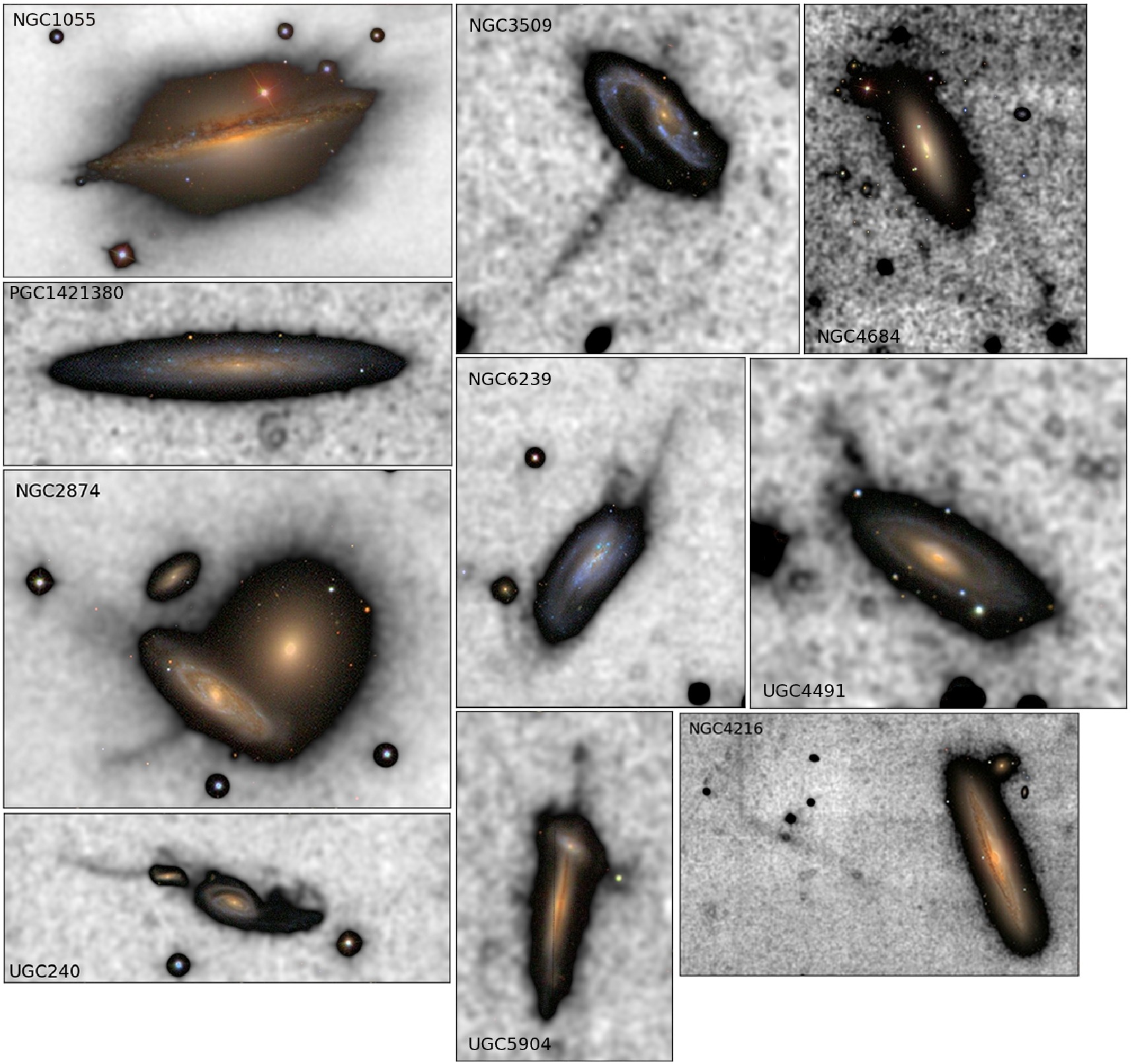}}
\caption{A selection of our found features overlayed with color images obtained by SDSS. NGC 1055 shows its large halo with spikes sticking out of it. NGC 3509 with its sword-like stream. NGC 4684 shows a large, loop-like feature. PGC 1421380 has a small, looped stream around it. NGC 6239 shows two large spikes sticking out. At UGC 5904, the interaction with a dwarf creates a stream which connects the galaxies. A complex interaction is exhibited at UGC 240. A very long stream is found at NGC 4216. A shorter, plume like feature is shown at UGC 4491. NGC 4216 and NGC 1055 are already known to have stream like features \citep{delgado2010}. Although some of the stream are very faint, it is clear that our method is at least able to show the existence of the feature around those galaxies.}
\end{figure*}

\textbf{I:} Features which show clear streams which are not the result of major mergers or other large scale interactions. Both were excluded by checking the vicinity of the galaxy for the presence of large galaxies. If none are found, large scale interactions and major mergers can be excluded. See fig. \ref{class1gall} for a complete collection of class 1 features (only electronically available).
\\

\textbf{II:} Features which seem to be connected to some disc disturbance like warping. Of course these could also be projections, but since disc disturbances cannot be excluded in these cases, they are not regarded as clear streams. See fig. \ref{class2gall} for a complete collection of class 2 features (only electronically available).
\\

\textbf{III}: Features which are possibly the result of (large) interacting galaxies and not between one large galaxy and its dwarf companion, where a minor merger origin cannot be certain. See fig. \ref{class3gall} for a complete collection of class 3 features (only electronically available).
\\

\textbf{IV}: Unclassified. Those would be features which make it hard to estimate what they really are, e.g. when it looks like a stream, but could also be very well an extension of a spiral arm. There is also the possibility that some situations overlap, e.g. there could be a tidally disrupted satellite leaving a spur, and being surrounded by multiple galaxies. If the remains of the dwarf would not clearly be visible, it would not be sure whether the detected feature would be a result from a tidal disruption of a dwarf galaxy or interaction within the galaxy group or even galactic cirrus, although this can be ruled out most of the time since it has a much more extended and diffuse structure than the found features. Furthermore it would be very unlikely that smaller, sharper parts of the cirrus would happen to be right in front of several small galaxies, distributed of a large portion of the sky, as it may be the case with single, larger galaxies like M81 \citep{cirrus}. See fig. \ref{class4gall} for a complete collection of class 4 features (only electronically available).

\begin{figure}[ht]
\centering
\includegraphics[width=\columnwidth]{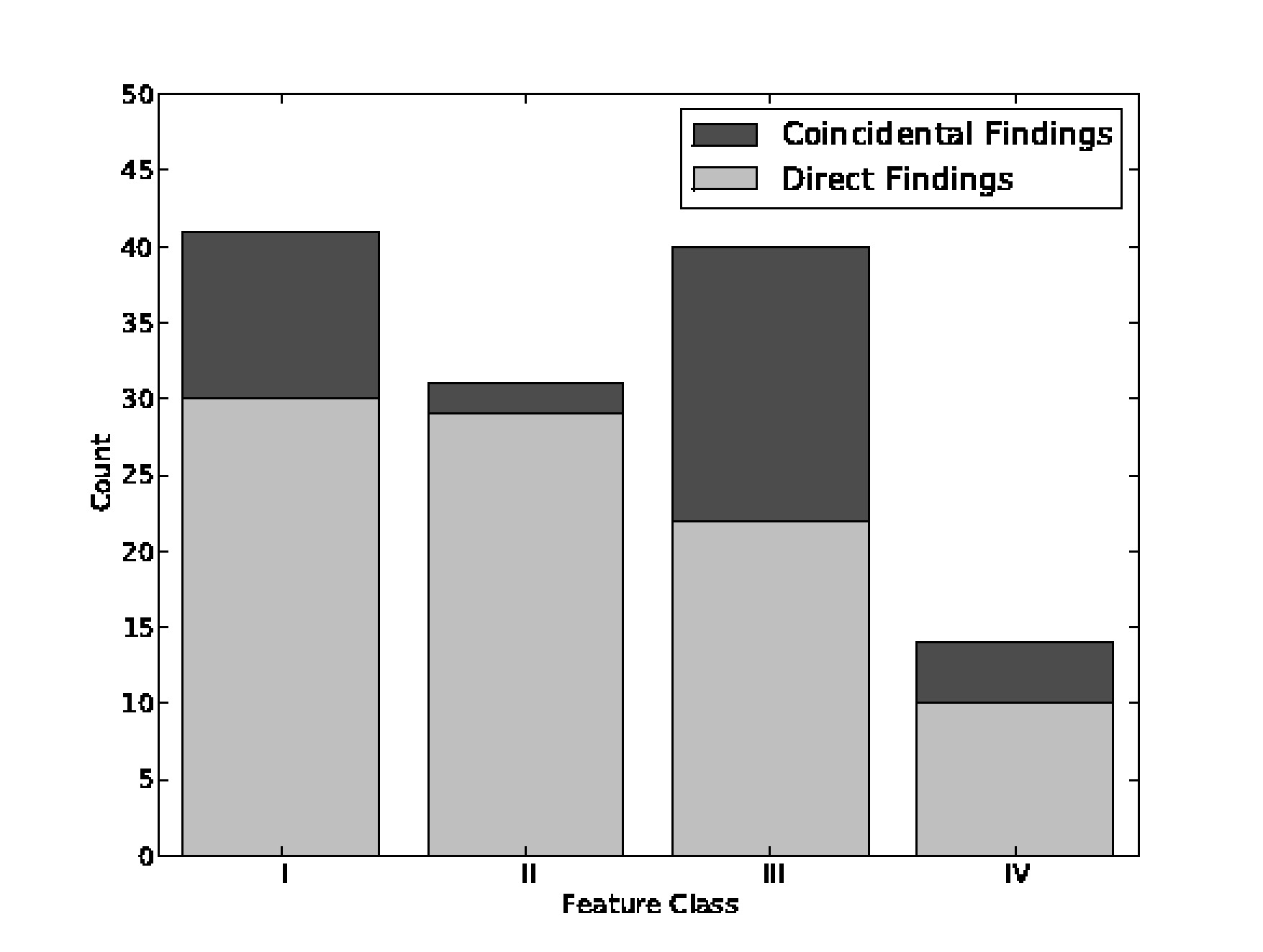}
\caption{Distribution of the four feature classes described in section \ref{label_res}, divided into findings from the input sample and coincidental findings. It shows a total of 41 galaxies in class I, 31 galaxies in class II, 40 galaxies in class III, and 14 galaxies in class IV,}
\label{classfreq}
\end{figure}

\subsection{Detection limit}
\begin{figure}[ht]
\centering
\includegraphics[width=0.9\columnwidth]{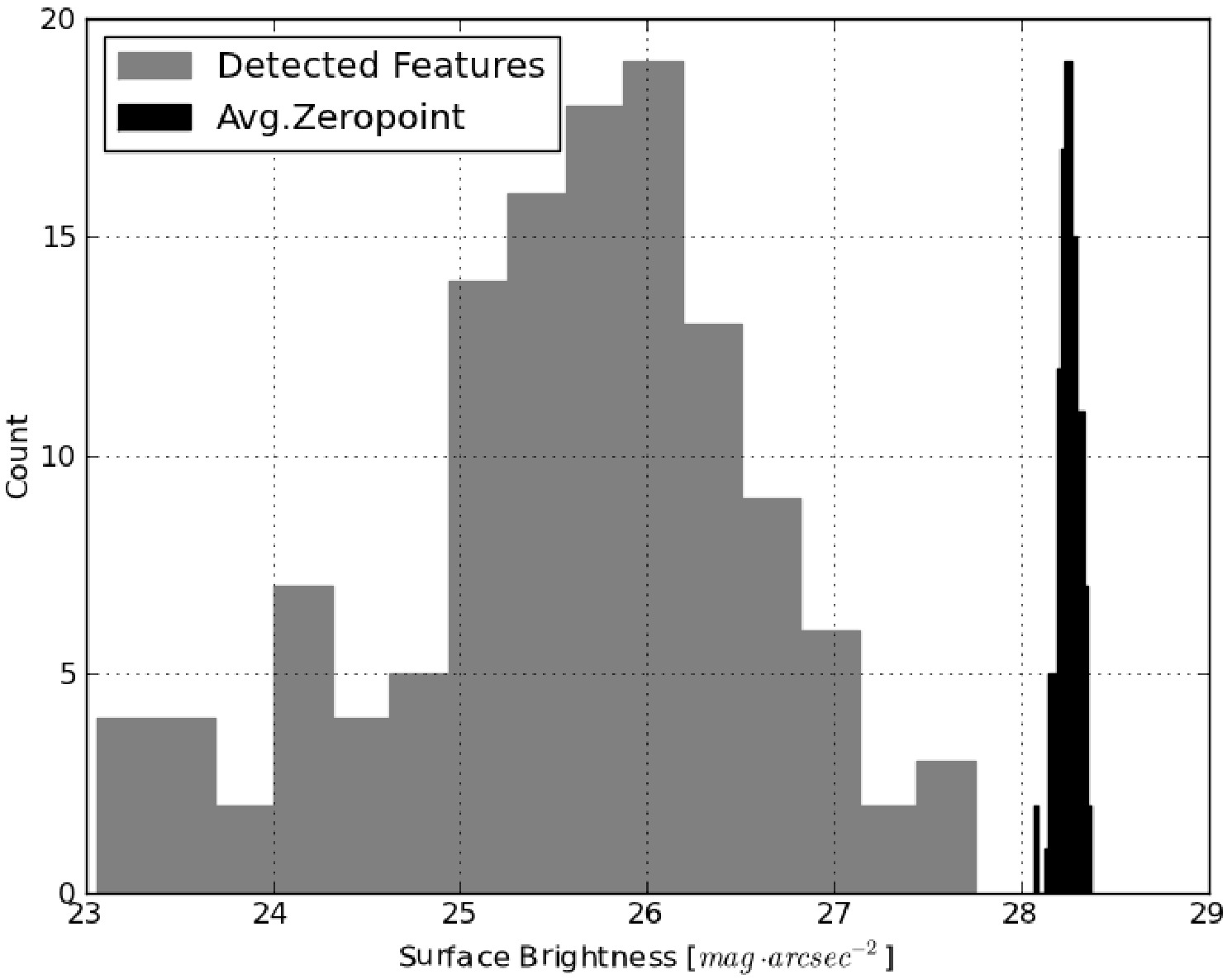}
\caption{Distribution of the the surface brightness of all the 126 detections, compared to the distribution of the average zero point of the surrounding fields. It shows that the detected surface brightnesses come very close to the detection limit of SDSS.}
\label{sbzeropoint}
\end{figure}

The analysis of the surface brightnesses of our detected features shows an estimate of the overall detection limit for faint features. A histogram of the distribution of the surface brightness has a clear peak at around 26 $mag ~ arcsec^{-2}$, see Fig. {sbzeropoint}. Comparing it to the distribution of the average zero point of the surrounding image fields, shows that the brightnesses are near to the average zero point. Fig. \ref{sbzeropoint} shows the histogram of the surface brightness of all 126 detections and the histogram of the averaged zero point of the surrounding fields. The steep decline in the right part of the histogram is due to the increased effect of the unflat background on low surface brightness features.

\subsection{Analysis examples using only SDSS Data}
In this section we will present two of the most striking findings of our work and a subsequent analysis of those objects to show the capabilities of this data. This analysis includes the length or extent of the found feature and if possible a time and mass calculation based on the equations given in \citep{johnston01}. For the length estimation, a general uncertainty of 2 arcseconds and a Hubble-constant of $h_0 = 73.2 km s^{-1} Mpc^{-1}$ \citep{wmap3y} was used.

\subsubsection{NGC 3509}
\label{ngc3509}
The first clear stream we found was at NGC 3509.

\begin{figure}[ht]
\centering
\includegraphics[width=\columnwidth]{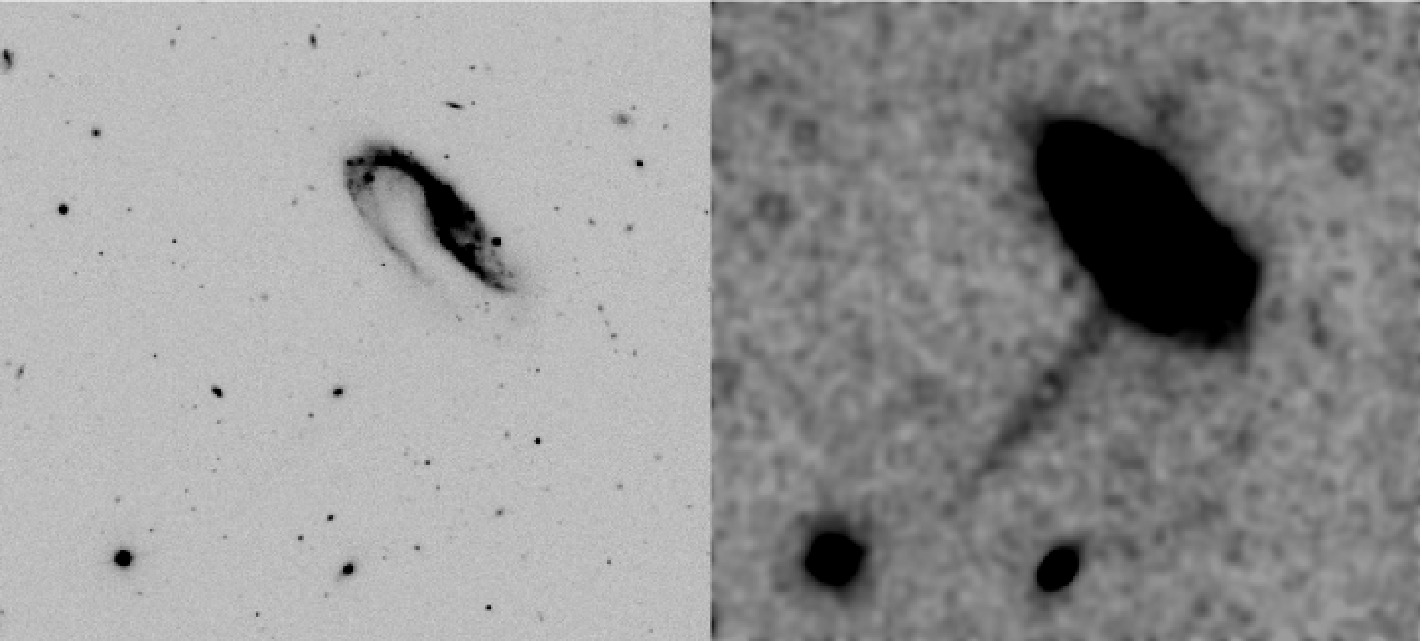}
\caption{Tidal stream associated with NGC 3509. The left image is an unprocessed image, the right one is processed as described in section \ref{processdata}.}
\label{3509}
\end{figure}

\citet{hsttoomre} analyzed this galaxy with the Hubble space telescope and concluded, that its shape is not the result of an ongoing major merger and mentioned that they have made deep images of the galaxy, and write about the small (dwarf) galaxy below NGC 3509 that "\textit{This possible companion also has faint low surface brightness features pointing both toward and away from NGC 3509}", though they have not analyzed it further. 
\\

The sword-like, low surface brightness feature, detected using our method, is shown in Fig. \ref{3509}. The stream was measured to have a length of 182 $\pm$ 2", which translates to a projected length of $ 92^{+5.3}_{-4.7} $ Kpc. This companion galaxy has, according to the SDSS spectral analysis, a very similar redshift as NGC 3509. It has a redshift of z = 0.02544 $\pm$ 0.000023\footnote{According to NED} and the companion has a redshift of z = 0.0257 $\pm$ 0.0001. These redshifts translate to velocities of 7530 km/s for the host galaxy and 7606 km/s for the companion, which yields a velocity difference of 76 km/s. The velocity difference between M31 and its satellite M32 is even higher than this (100 km/s). This definitely means that both galaxies are very close together and that the companion is highly likely bound to the host galaxy. Furthermore, the dwarf galaxy has an absolute magnitude of -16.62 in the B band, a B - V color of 0.77 and is about 2 Kpc in diameter. Those properties would suggest, that this dwarf galaxy is a dE type dwarf galaxy. 

\begin{figure}[ht]
\centering
\includegraphics[width=\columnwidth]{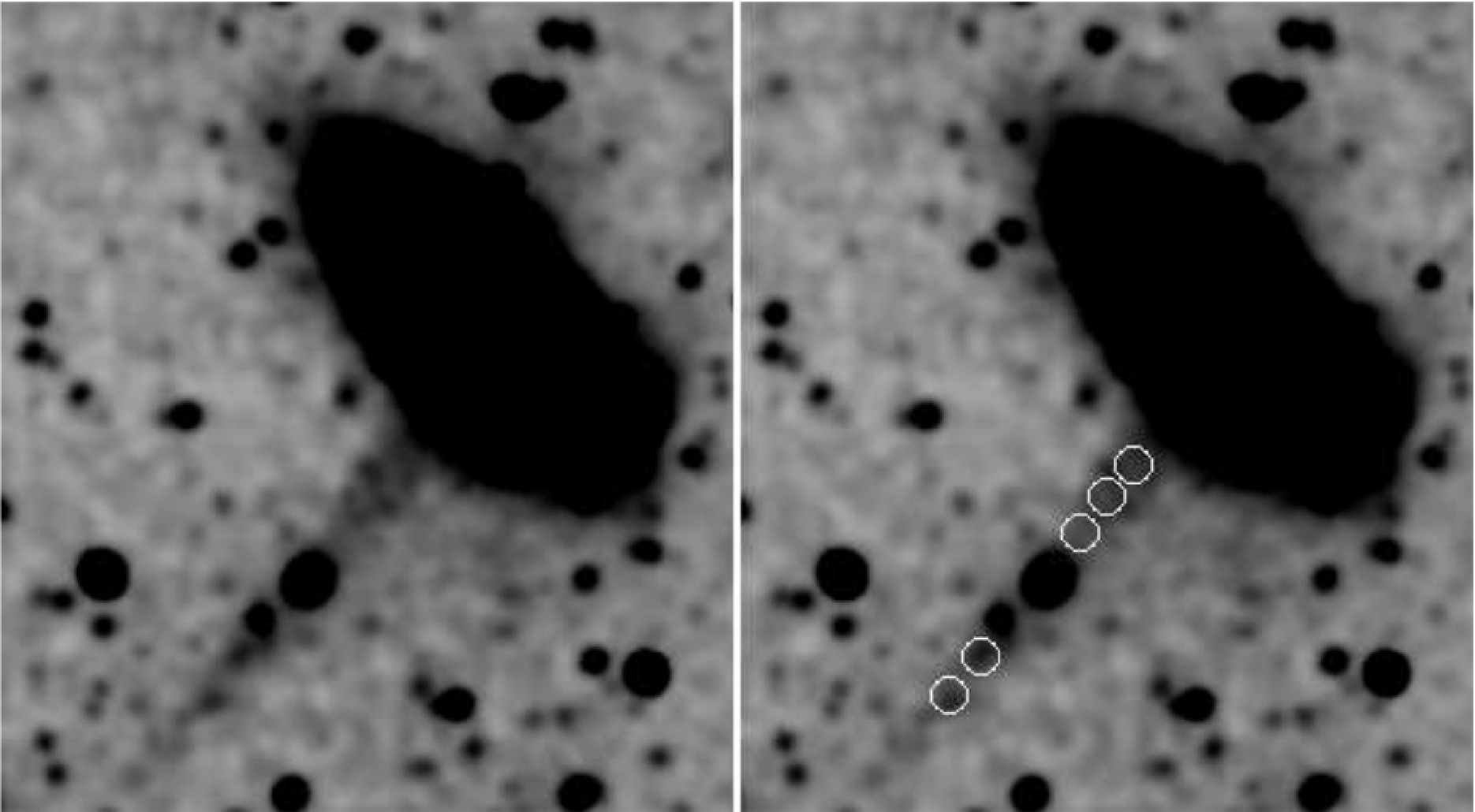}
\caption{The stream at NGC 3509 from the processed SDSS data, and on the right, the same image with overlaid photometry points.}
\label{3509regions}
\end{figure}

Since the stream is clearly visible in the SDSS image, it was the first stream to be measured photometrically. In all five bands, five photometry points (Fig. \ref{3509regions} shows their position on the stream) were laid over the stream, the average flux in them was measured and converted to surface brightness.
\begin{table}[ht]
\caption{Surface brightnesses for the five photometry points on the stream at NGC 3509. Filters u' and z' are not sensitive enough to show the stream.}             
\label{3509table}     
\centering
\begin{tabular}{c c c c}
\hline
Filter & Integrated Magnitude & \textbf{$\mu$} & error\\
\hline
u & - & \textbf{-} & -\\
g & 20.61 & \textbf{27.21} & 0.15\\
r & 20.08 & \textbf{26.92} & 0.13\\
i & 21.08 & \textbf{26.84} & 0.29\\
z & - & \textbf{-} & -\\
\hline                                   
\end{tabular}
\end{table}

Of course these magnitudes are to be treated carefully, since they were obtained from noisy images. But the values are similar to values obtained by Mart\'{i}nez-Delgado et al. for NGC 5907 \citep{delgado2008} and NGC 4013 \citep{delgado2009}. The magnitudes of the stream indicate, that the stream is at its brightest in the i' band. This indicates, that there is no significant (recent) star forming going on in this stream, or else the stream would be much bluer and either the g' or r' band would be brighter. We also calculated, that the stream has a B - V color of 0.71, which is close to the B - V color of the dwarf. The discrepancy is highly likely due to the low flux levels of the stream. But there is no doubt, that this feature originates from the dwarf galaxy.
To get a rough estimate of the stellar population of the dwarf and the stream we compared the colors with predictions of \citet{maraston1998}. The predictions are based on a simple stellar population model (SSP), using a single star formation burst at a given time, a Salpeter IMF and a red branch morphology of the horizontal branch. The tables used for the comparison give information on the metallicity, age and the integrated magnitudes in SDSS ugriz filters. A good match for the observed colors yields a population with the parameters given in Tab. \ref{param_table}.

\begin{table}[ht] 
\centering
\begin{tabular}{|c c c c|}
\hline
At & Z/H & t [Gyr]& B - V\\
\hline
Dwarf & -1.35 & 9 & 0.77\\
Stream & -1.35 & 5 & 0.7\\
\hline
\end{tabular}
\caption{Parameters and calculated B - V of the SSP's that best fit our observed colors for the dwarf galaxy and stream at NGC 3509.}             
\label{param_table}
\end{table}

This model would suggest that the dwarf is a primordial dwarf which had its last major starburst about 9 Gyrs ago.

\subsubsection{Age of the Stream and Mass of the Dwarf}
\citep{johnston01} derived an analytical description of a dwarf being tidally destroyed by its host galaxy, and tested it with N-body simulations on found streams like NGC 5907. In their work they derived approximations for the mass of the progenitor dwarf galaxy, and the time since its disruption. The approximation for the initial mass of the dwarf is
\begin{eqnarray}
\label{dwarfmass}
m \sim 10^{11} \left(\frac{w}{R}\right)^3\left(\frac{R_p}{10~kpc}\right)\left(\frac{v_{circ}}{200~km s^{-1}}\right)^2 M_{\odot}
\end{eqnarray}
and the approximation for the time since the disruption is
\begin{eqnarray}
\label{disruptiontime}
t \sim 0.01\Psi \left(\frac{R}{w}\right)\left(\frac{R_{circ}}{10~kpc}\right)\left(\frac{200~km s^{-1}}{v_{circ}}\right) Gyr
\end{eqnarray}
The variables are $w$, the widest width of the stream at radius $R$, $R_p$ is the pericentric distance of the streams orbit, $\Psi$ the angular length in radiants, $R_{circ}$ is the radius of the circular orbit with the same energy as the true orbit, and $v_{circ}$ is the maximal circular velocity of the host galaxy. $W$, $\Psi$, $R$ and $R_p$ can be measured directly in the images, $v_{circ}$ can be measured using radio telescopes or looked up, e.g. in the HyperLeda database \citep{hyperleda}. $R_{circ}$ however cannot be measured directly, so in this case, it was approximated as being half way between the pericentric and apocentric distance.
When measuring the radii, one has to take into account that the host galaxy and with that the stream is inclined at a specific angle. The "true" radius R for example, is then calculated as $R = R_{obs} / cos(i)$, where i is the inclination angle and $R_{obs}$ the radius measured directly in the image and converted to Kpc using the pixel scale of the image and the distance to the galaxy.

This calculation has been adopted for the galaxies in this section to calculate the age of the stream and the mass of the dwarf. For NGC 3509, the measured values are:
$w = 9.1~kpc$, $R_p = 61.66~kpc$, $R = 205.69~kpc$\footnote{Both radii were calculated using an inclination angle of $63.62 \deg$}, $\Psi = \frac{1}{2}\pi = 1.57$, $R_{circ} = 72.02~kpc$ and $v_{circ} = 198~km~s^{-1}~$\footnote{Taken from the HyperLeda Database}. There are however some difficulties with the measurements of this galaxy. One difficulty is the most obvious one; NGC 3509 is a rather irregularly shaped galaxy, which makes the determination of the inclination angle a bit uncertain. Another problem is, that the stream of NGC 5907 seems to get wider with projected distance from the center of the main galaxy. Here however, the stream seems to get wider the closer it is projected to the center of NGC 3509, which means that $R$ and $R_p$ are roughly the same. Using these radii, the mass is calculated to $m = 1.94 \cdot 10^9 M_{\odot}$ and the time to t = 0.77 Gyr. These values are very similar to those for the stream of NGC 5907. Another small difficulty is, that the stream of NGC 3509 seems to be a straight line, so its angular length cannot be determined accurately. For this analysis, it was assumed to have an angular length of $\frac{1}{2}\pi$, since such a shape would produce a straight line when looked from above. Because the dwarf is slightly further away than the host galaxy and appears to be in the middle of the stream, and the stream gets narrower the with distance to the main galaxy, it is a reasonable assumption, that the stream "creates" at least a quarter of a circle. Of course it could be a bit longer, but not visible due to brightness limitations, or a bit shorter, so a quarter of a circle should be a good approximation.

\subsubsection{NGC 7711}
Another interesting finding is the S0\footnote{according to NED} galaxy NGC 7711 which lies at the J2000 coordinates RA = 23h 35m 39.4s, DEC = +15$^\circ$ 17m 7s, at a distance of $55.5^{+2.9}_{-2.6}$ Mpc. This seemingly isolated galaxy seems to be not mentioned in any obvious way in literature. We found in the stacked image, that there is an extension of the disc with a gap in the middle. It looks to us like this extensions are in fact tidal streams of a smaller dwarf galaxy which completed at least two orbits around NGC 7711. This is supported by a bright spot which is very close to the disk of NGC 7711, so close that it could be interpreted as part of the disk, which would then appear very asymmetrically. The outer stream has an extent of $72" \pm 2"$ from the d25 line, which translates to about 19.4 kpc.
\begin{figure}[ht]
\centering
\includegraphics[width=\columnwidth]{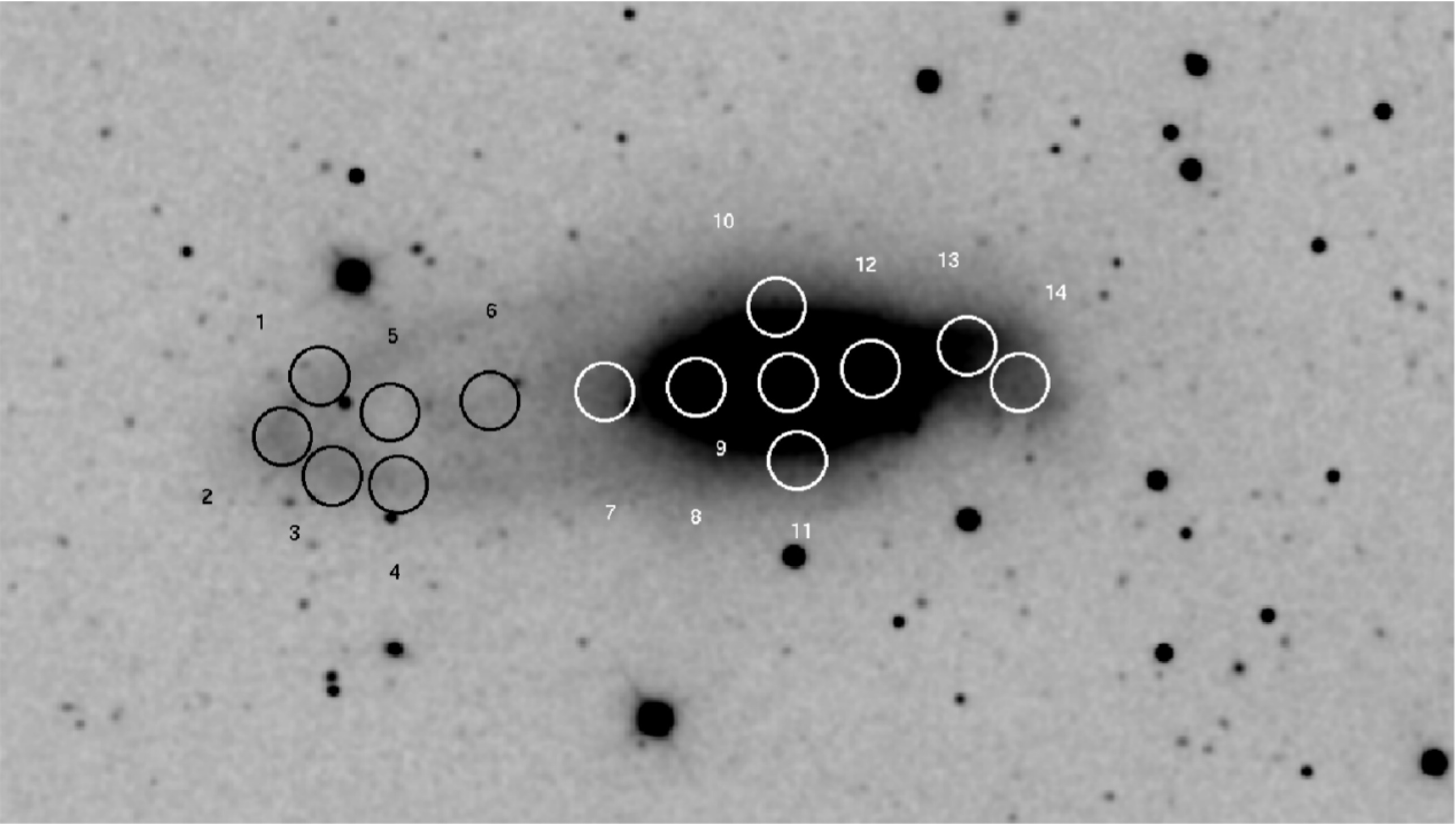}
\caption{SDSS image of the stream surrounding NGC 7711 with a FOV of 6 by 3 arcminutes. Overlaid are 14 photometry points along the assumed arc and on the galaxy itself.}
\label{ngc7711regions}
\end{figure}

Using the IRAF task phot, the surface brightness in Pogson magnitudes of each point was determined. The first 6 points were laid onto the stream, where point 5 serves a test to see if there is indeed a gap between the inner and outer stream. Points 7 to 14 are laid on the galaxy and the lump on its right side to check its surface brightness and later the color of the galaxy and stream.

\begin{table}[htbp]
\caption{Surface brightnesses of the photometry points}
\centering
\begin{tabular}{c c c c c c}
\hline
\textbf{point} & \textbf{u} & \textbf{g} & \textbf{r} & \textbf{i} & \textbf{z} \\ \hline
1 & 26.49 & 25.09 & 24.38 & 23.95 & 23.71 \\
2 & 26.46 & 24.64 & 23.9 & 23.64 & 23.26 \\ 
3 & 26.31 & 24.81 & 24.05 & 23.71 & 23.32 \\ 
4 & 26.39 & 25.00 & 24.26 & 23.98 & 23.61 \\ 
5 & 27.12 & 25.36 & 24.78 & 24.33 & 23.78 \\ 
6 & 26.64 & 24.97 & 24.23 & 23.84 & 23.37 \\ 
7 & 25.7 & 24.06 & 23.26 & 22.83 & 22.48 \\ 
8 & 23.92 & 22.23 & 21.46 & 21.04 & 20.77 \\ 
9 & 21.6 & 19.72 & 18.83 & 18.38 & 18.04 \\ 
10 & 25.09 & 23.32 & 22.52 & 22.03 & 21.84 \\ 
11 & 25.13 & 23.52 & 22.73 & 22.2 & 21.99 \\ 
12 & 23.94 & 22.17 & 21.37 & 20.92 & 20.65 \\ 
13 & 25.22 & 23.35 & 22.58 & 22.17 & 21.88 \\ 
14 & 25.55 & 23.93 & 23.18 & 22.76 & 22.45 \\
\hline 
\end{tabular}
\caption{Surface brightness in each SDSS band for every photometry point on NGC 7711, the stream and the suspected dwarf.}
\label{7711phot}
\end{table}
As Tab. \ref{7711phot} shows, point 5 has the lowest surface brightness of all points, which clearly means that this in fact is a gap between the inner and outer stream. Point 9 has the highest value, which is what one would expect, since point 9 lies on the center of the galaxy. This is by now, the most clearest and brightest of all detected streams around a galaxy. With these calculated magnitudes, it is possible to create color-color diagrams for each point in order to see if it reveals some more information. The colors used for the diagram were g' - r' and u' - g'. The errors in the determination of the magnitudes are very low due to the high fluxes. The highest error has point 5 in the u' band, with an error of $\sim$ 0.4 mag. This only would affect the u' - g' color, and not the position on the g' - r' axis. These errors however are calculated for the integral magnitude in the 20 pixel radius aperture, which ignores possible correlations between those pixels.
\begin{figure}[ht]
\centering
\includegraphics[width=\columnwidth]{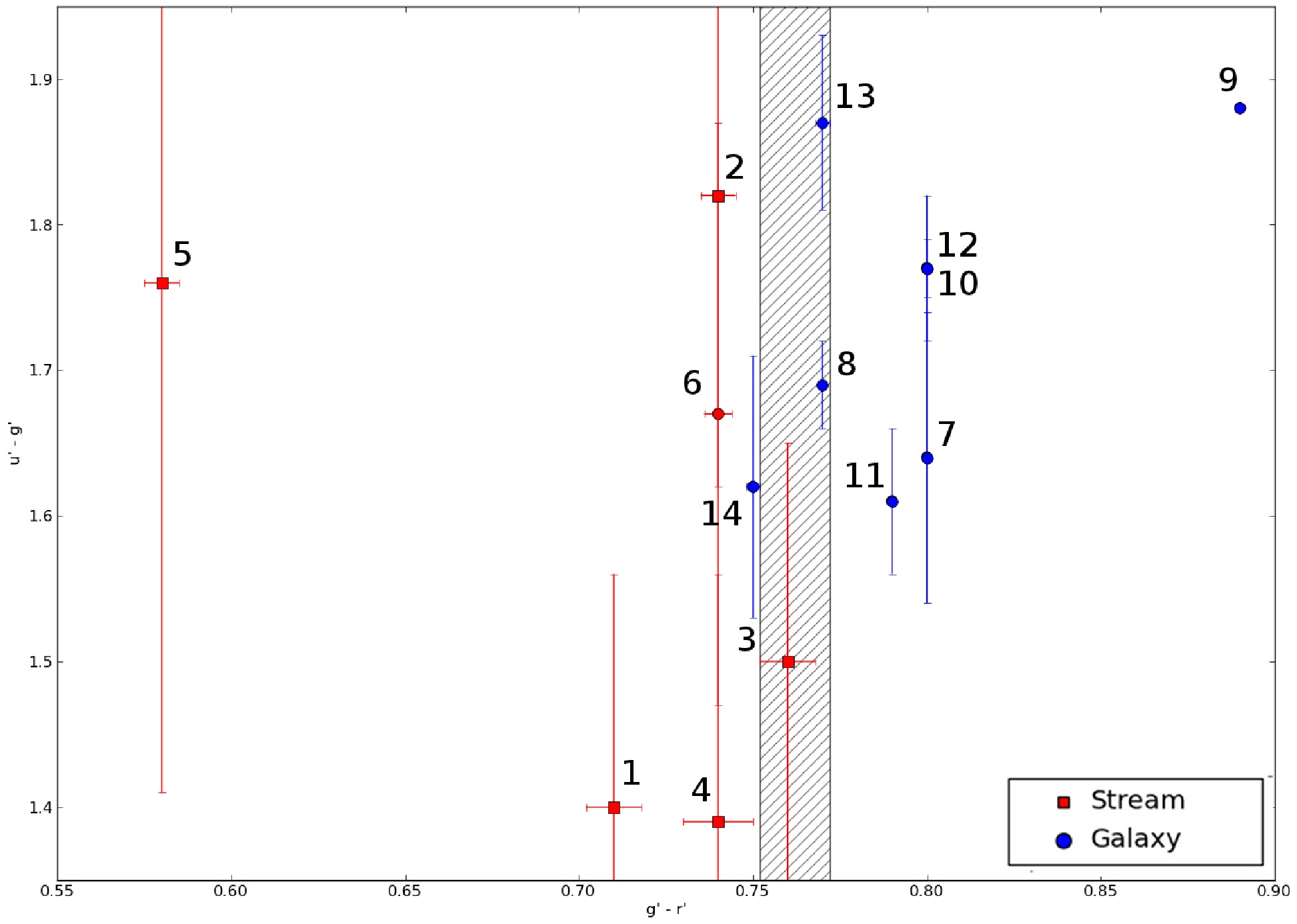}
\caption{Color-color diagram of the 14 photometry points for NGC 7711. Blue points belong to NGC 7711 and red points to the stream surrounding it. Shaded area indicates the area where the g'-r' distinction between galaxy and stream lies.}
\label{ngc7711color}
\end{figure}

The blue points in Fig. \ref{ngc7711color} are points 7 to 14, which lie on the galaxy, and the red points are 1 to 6, the points on the stream. The shaded area on the diagram indicates the area where the distinction in the g' - r' color between the galaxy and the stream was made. The numbers beside the points are the same numbers that each point on Fig. \ref{ngc7711regions} has. As one can see, point 14, which lies on the lump right to the galaxy, seems to be in the same g' - r' population as the points on the stream, instead of the higher g' - r' population of the galaxy's disc. This implies, that this lump really is the dwarf which orbits the main galaxy, or at least, the remains of it. Point 13, however, which was also put on the lump, but a bit closer to the disc, seems to be in the galaxy population of the g' - r' color, although it is very close to the line. It is possible that the color at this point is "contaminated" with light from the disc. One could argue, that point 3 might belong to the galaxy population because it is very close to the "border", but looking at the image, point 3 clearly belongs to the stream.

To get a rough stellar population, we compared it again to the models of \citep{maraston1998}. An SSP was applied again. We averaged the integral magnitudes of the disc, the stream and the possible dwarf, and then searched matching pairs in the SSP models. Tab. \ref{avg_table} shows the average integral magnitudes of the NGC 7711 disc, the dwarf and the stream. We included the disc of NGC 7711 to see if the possible dwarf on the right side has a different composition or if it is just part of the disc.

\begin{table}[ht] 
\centering

\begin{tabular}{|c c c|}
\hline
At &g'-r'&u'-g'\\
\hline
Disc & 0.86 & 1.85\\
Stream & 0.74 & 1.56\\
Dwarf & 0.76 & 1.77\\
\hline
\end{tabular}
\caption{Average g-r and u-g values for the disc of NGC7711, the stream and the dwarf on the right.}             
\label{avg_table}
\end{table}

\begin{table}[ht] 
\centering
\begin{tabular}{|c c c c c c c c|}
\hline
At & Z/H & t [Gyr]& u & g & r &g'-r'&u'-g'\\
\hline
Disc & 0.35 & 5 & 8.98 & 7.11 & 6.25 & 0.86 & 1.87\\
Stream & 0 & 3 & 7.92 & 6.32 & 5.58  & 0.74 & 1.6\\
Dwarf & 0 & 4 & 8.22 & 6.59 & 5.83  & 0.76 & 1.63\\
\hline
\end{tabular}
\caption{Parameters and calculated g-r, u-g of the SSP's that best fit our observed colors for the disc of NGC 7711, the stream and the possible dwarf galaxy}             
\label{7711ssp}
\end{table}

The SSP model, seen in Tab. \ref{7711ssp},  shows the best match of the g'-r' and u'-g' pair for the disc of NGC 7711 and the worst match for the possible dwarf. That might be hint for a more complex population in the supposed dwarf. The matches for the dwarf and the stream shows the same metallicity and a similar age, which indicates that the stream originates from the dwarf. The higher metallicity of the disc is expected, since a minor merger should not lower the SFR, but rather increase it locally. The ages suggest that the dwarf is not a primordial dwarf as the one at NGC 3509, but possible already the result of a merger.

\section{Conclusions and future work}
The occurrence of faint features, which point to gravitational interaction of galaxies with smaller dwarf galaxies seems to be at a significant level. 19\% of the galaxies in our input sample showed low surface brightness features connected, or near the disc, at least 6\% showed distinct stream like features. Due to uncertainties with the background, which limits our detection capabilities, this should be a lower limit on the frequency of such features. The streams that we found seem to be in different stages of accretion; we found relatively short streams, like at NGC 3221, medium sized streams like at NGC 4684 and even a stream, which seems to have already completed two orbits around NGC 7711. This indicates that faint features like tidal streams seem to be common and play a role in the evolution of galaxies and are still an ongoing process, at least in the local universe. A detailed statistical analysis will be presented in paper two. The measured surface brightness of the stream near NGC 3509 lies between 26.8 and 27.3 $mag ~ arcsec^{-2}$, which is similar to the surface brightnesses measured for similar features at other galaxies by \citet{delgado2008}. The stream at NGC 7711 is even brighter, ranging between around 23.5 and 26.5 $mag  arcsec^{-2}$, making it one of the brightest stream that has been discovered yet. 

In this paper we showed that SDSS data are, despite the short integration time, capable of showing almost every faint feature which was previously shown by others like e.g. by \citep{delgado2010}, proving that searching for faint features with the Sloan Digital Sky Survey is possible. We also found several serendipitous features near other galaxies in our fields, which implies that searching for more galaxies from a much larger input sample with less rigorous selection criteria should yield many more faint features. This led us to the next step: At this time, we are processing and analyzing every galaxy from the RC3 catalog which was imaged by SDSS. We expect to find many more streams out of this input sample and to be able to derive more accurate statistics, which will allow us to search for correlations with global parameters of the galaxies and their environment.

\begin{acknowledgements}
~\\
    We thank David Martinéz-Delgado for his permission to use his images of NGC 4013 and NGC 5907.
\\
    We thank David Malin and Steve Mandel for their permission to use the images of M104 and NGC 3628.
\\
    We thank the anonymous referee for his constructive and helpful comments which helped us improve our paper.
\\
    This research made use of Montage, funded by the National Aeronautics and Space Administration's Earth Science Technology Office, Computation Technologies Project, under Cooperative Agreement Number NCC5-626 between NASA and the California Institute of Technology. Montage is maintained by the NASA/IPAC Infrared Science Archive.
\\
    We acknowledge the usage of the HyperLeda database (http://leda.univ-lyon1.fr).
\\
    This research has made use of the NASA/IPAC Extragalactic Database (NED) which is operated by the Jet Propulsion Laboratory, California Institute of Technology, under contract with the National Aeronautics and Space Administration.
\\
    Funding for the SDSS and SDSS-II has been provided by the Alfred P. Sloan Foundation, the Participating Institutions, the National Science Foundation, the U.S. Department of Energy, the National Aeronautics and Space Administration, the Japanese Monbukagakusho, the Max Planck Society, and the Higher Education Funding Council for England. The SDSS Web Site is http://www.sdss.org/.
\\
    The SDSS is managed by the Astrophysical Research Consortium for the Participating Institutions. The Participating Institutions are the American Museum of Natural History, Astrophysical Institute Potsdam, University of Basel, University of Cambridge, Case Western Reserve University, University of Chicago, Drexel University, Fermilab, the Institute for Advanced Study, the Japan Participation Group, Johns Hopkins University, the Joint Institute for Nuclear Astrophysics, the Kavli Institute for Particle Astrophysics and Cosmology, the Korean Scientist Group, the Chinese Academy of Sciences (LAMOST), Los Alamos National Laboratory, the Max-Planck-Institute for Astronomy (MPIA), the Max-Planck-Institute for Astrophysics (MPA), New Mexico State University, Ohio State University, University of Pittsburgh, University of Portsmouth, Princeton University, the United States Naval Observatory, and the University of Washington.
\\
    Based partly  on observations made with the NASA/ESA Hubble Space Telescope, obtained from the data archive at the Space Telescope Science Institute. STScI is operated by the Association of Universities for Research in Astronomy, Inc. under NASA contract NAS 5-26555.

\end{acknowledgements}

\bibliographystyle{bibtex/aa} 
\bibliography{bibtex/biblio}

\Online

\begin{figure*}[ht]
\centering
\includegraphics[width=\textwidth]{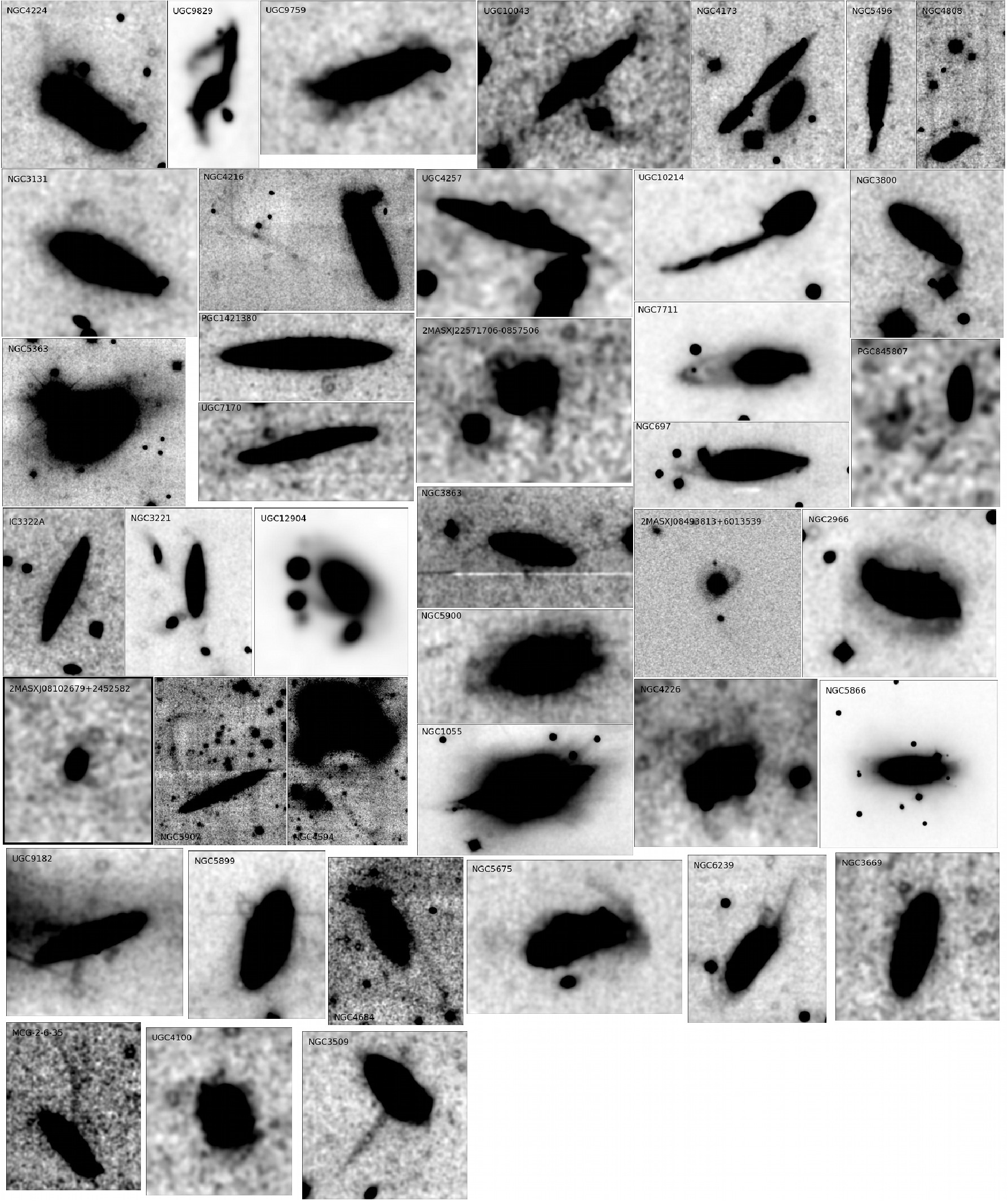}
\caption{Compilation of thumbnails of the 41 galaxies with class I Features. 2MASX J081026.79+245258.2, 2MASX J084938.13+601353.9, 2MASX J225717.06-085750.6, IC 3322A, MCG 2-6-35, NGC 0697, NGC 1055, NGC 2966, NGC 3131, NGC 3221, NGC 3509, NGC 3800, NGC 3863, NGC 4173, NGC 4216, NGC 4224, NGC 4226, NGC 4594, NGC 4684, NGC 4808, NGC 5363, NGC 5496, NGC 5675, NGC 5866, NGC 5899, NGC 5900, NGC 5907,  NGC 6239, NGC 7711, UGC 4257, PGC 845807, PGC 1421380, UGC 4100, UGC 7170, UGC 9182, UGC 9759, UGC 9829, UGC 10043, UGC 10214, UGC 12904.}
\label{class1gall}
\end{figure*}

\begin{figure*}[ht]
\centering
\includegraphics[width=\textwidth]{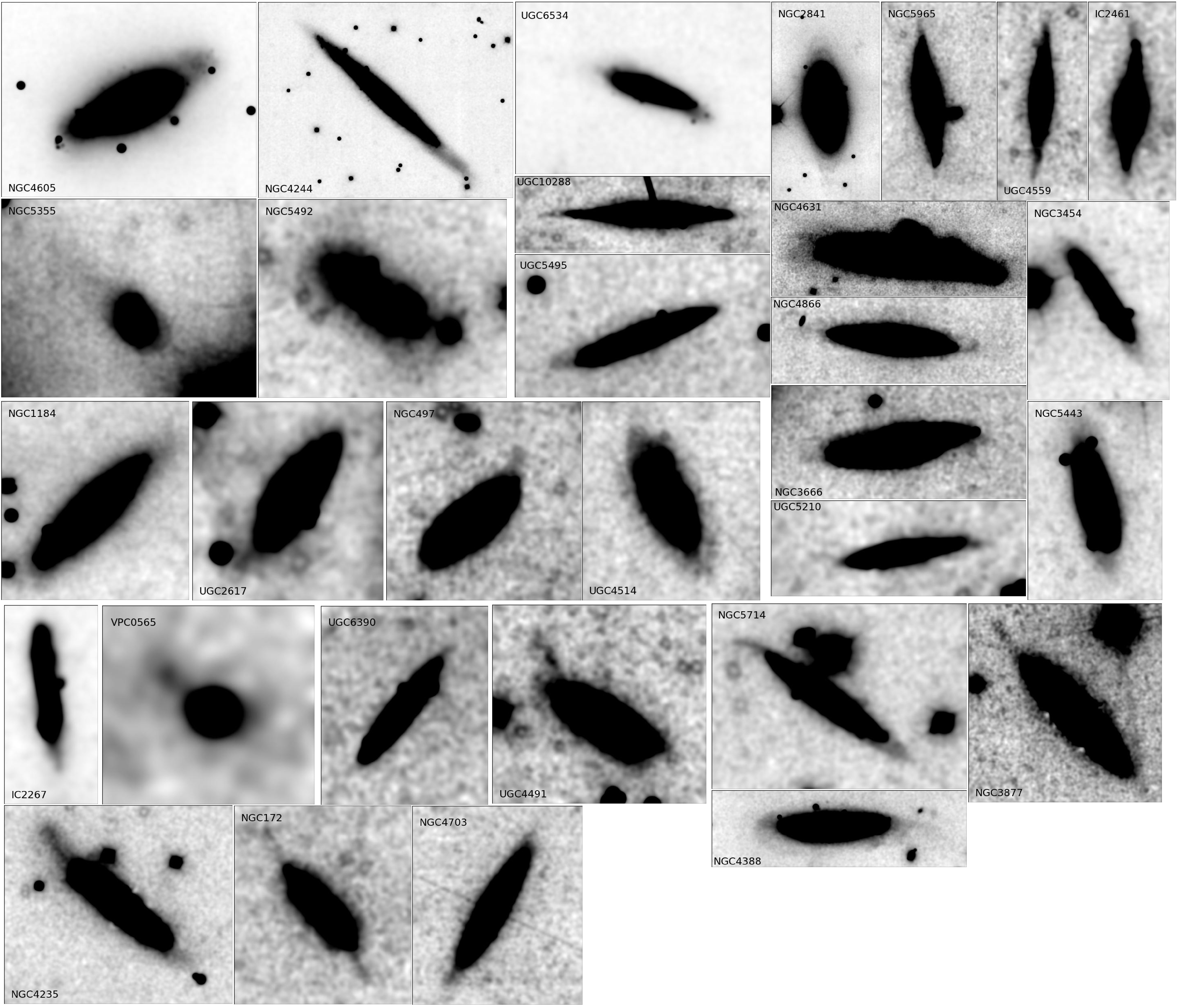}
\caption{Compilation of thumbnails of the 31 galaxies with class II Features. IC 2267, IC 2461, NGC 0172, NGC 0497, NGC 1184, NGC 2841, NGC 3454, NGC 3666, NGC 3877, NGC 4235, NGC 4244, NGC 4388, NGC 4605, NGC 4631, NGC 4703, NGC 4866, NGC 5355, NGC 5443, NGC 5492, NGC 5714, NGC 5965, UGC 2617, UGC 4491, UGC 4514, UGC 4559, UGC 5210, UGC 5495, UGC 6390, UGC 6534, UGC 10288, VPC 0565.}
\label{class2gall}
\end{figure*}

\begin{figure*}[ht]
\centering
\includegraphics[width=\textwidth]{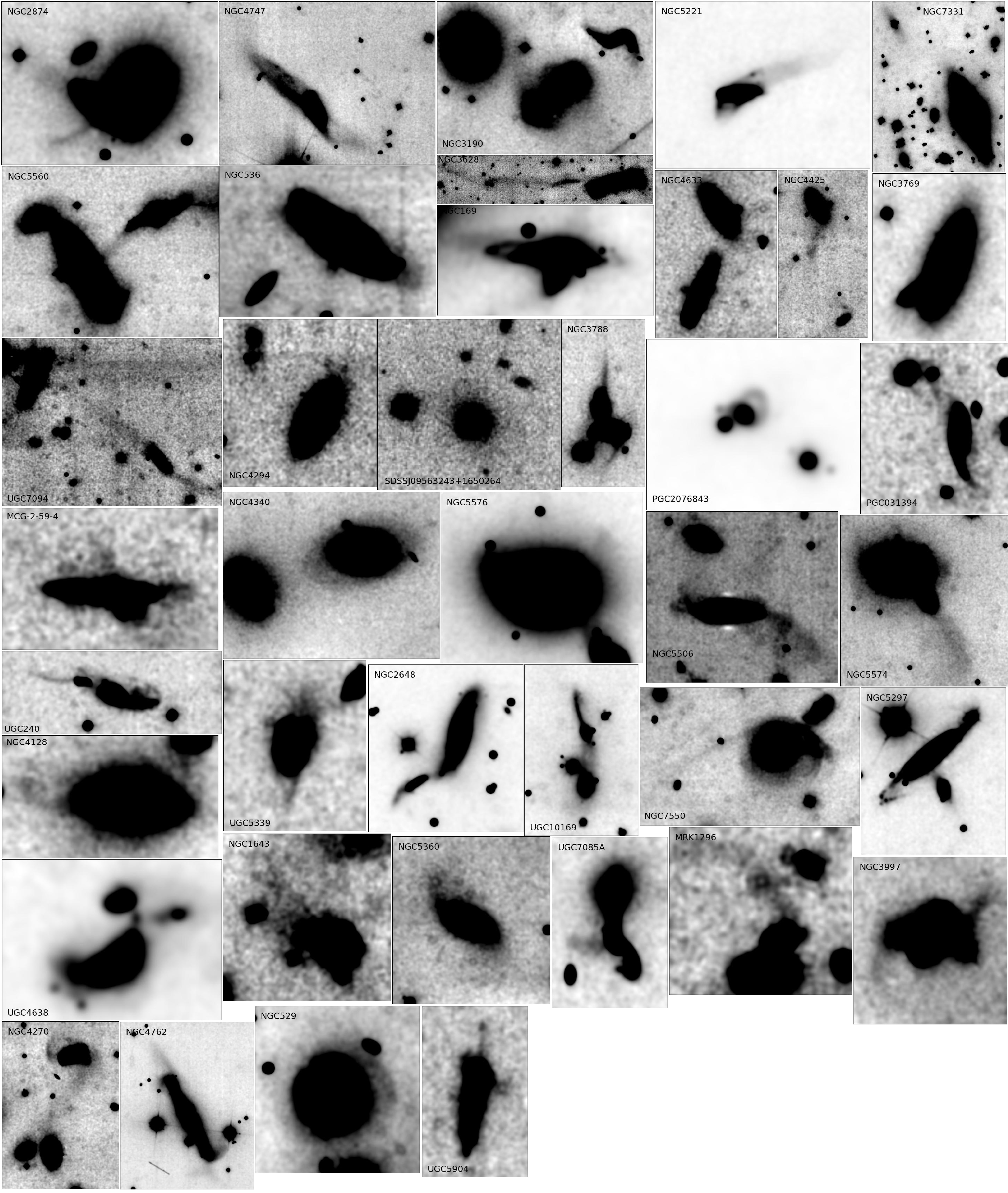}
\caption{Compilation of thumbnails of the 40 galaxies with class III Features. MCG 2-59-4, MRK 1296, NGC 0169, NGC 0529, NGC 0536, NGC 1643, NGC 2648, NGC 2874, NGC 3190, NGC 3628, NGC 3769, NGC 3788, NGC 3997, NGC 4128, NGC 4270, NGC 4294, NGC 4340, NGC 4452, NGC 4633, NGC 4747, NGC 4762, NGC 5221, NGC 5297, NGC 5360, NGC 5506, NGC 5560, NGC 5574, NGC 5576, NGC 7331, NGC 7550, PGC 031394, PGC 2076843, SDSS J095632.43+165026.4, UGC 0240, UGC 4638, UGC 5399, UGC 5904, UGC 7085A, UGC 7094, UGC 10169, }
\label{class3gall}
\end{figure*}

\begin{figure*}[ht]
\centering
\includegraphics[width=\textwidth]{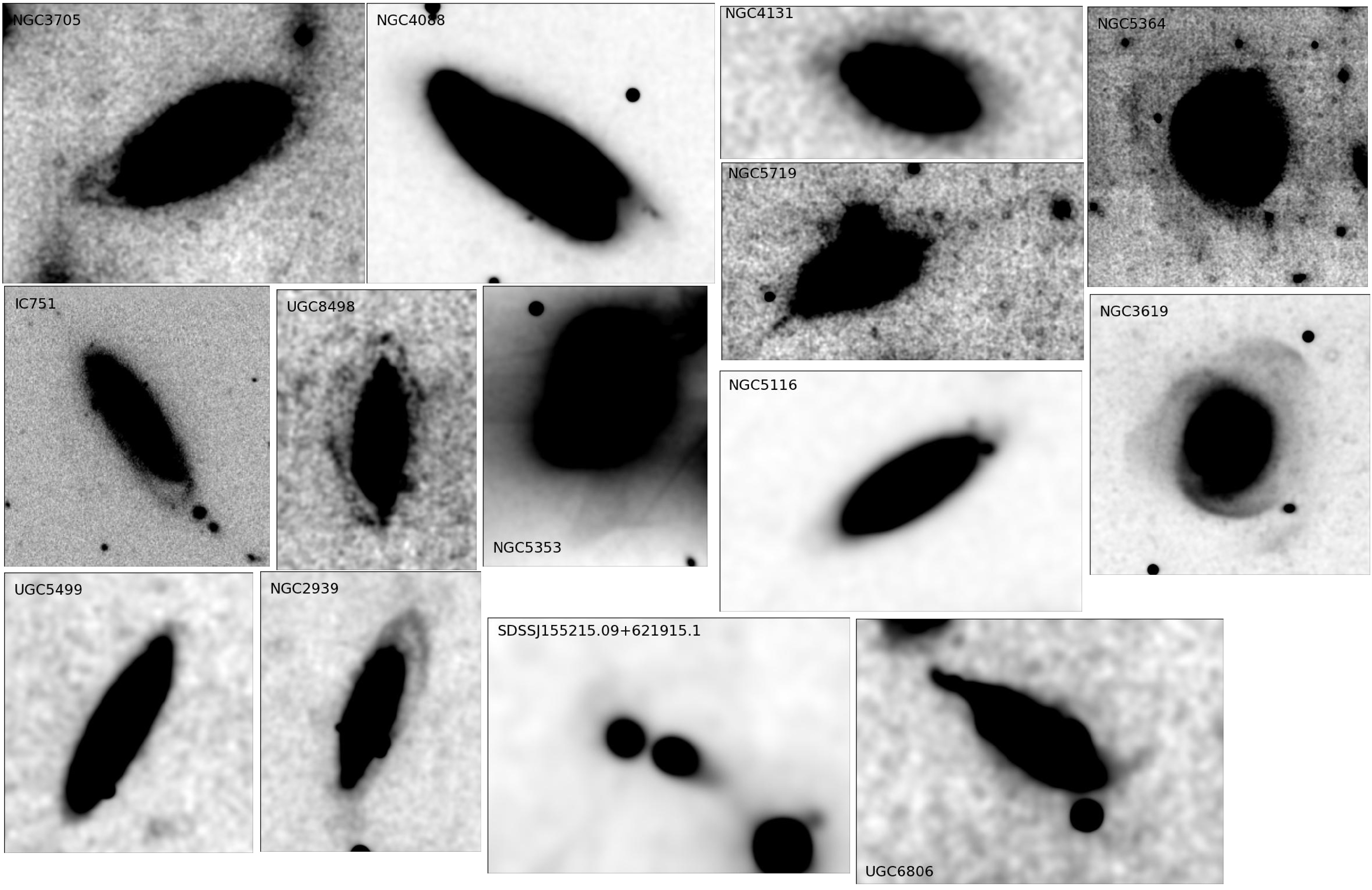}
\caption{Compilation of thumbnails of the 14 galaxies with class IV Features. IC 751, NGC 2939, NGC 3619, NGC 3705, NGC 4088, NGC 4131, NGC 5116, NGC 5353, NGC 5364, NGC 5791, SDSS J155215.09+621915.1 UGC 5499, UGC 6806, UGC 8498.}
\label{class4gall}
\end{figure*}

\end{document}